\documentclass[12pt]{article}
\usepackage{graphicx,tabularx,epsfig}
\usepackage{color}
\usepackage{enumitem}
\usepackage{caption}
\usepackage{subcaption}
\usepackage{amsmath}
\usepackage{amssymb}
\usepackage{amsthm}
\usepackage{physics}
\usepackage{dsfont, mathtools}
\usepackage{psvectorian}
\usepackage{blkarray}
\usepackage{bm}
\usepackage{url}
\usepackage{setspace}

\usepackage{xcolor}
\colorlet{linkequation}{blue}

\usepackage[colorlinks = true,
            linkcolor = blue,
            urlcolor  = blue,
            citecolor = blue,
            anchorcolor = blue]{hyperref}

\usepackage{floatrow}

\newlength{\abstractwidth}
\renewcommand{\thefootnote}{\fnsymbol{footnote}}
\renewcommand{\thanks}[1]{\footnote{#1}} 
\newcommand{\starttext}{
\setcounter{footnote}{0}
\renewcommand{\thefootnote}{\arabic{footnote}}}

\makeatletter
\g@addto@macro\normalsize{%
  \setlength\abovedisplayskip{10pt}
  \setlength\belowdisplayskip{20pt}
  \setlength\abovedisplayshortskip{10pt}
  \setlength\belowdisplayshortskip{20pt}
}
\makeatother

\usepackage{color}

\renewcommand{\title}[1]{\vbox{\center\LARGE{#1}}\vspace{5mm}}
\renewcommand{\author}[1]{\vbox{\center#1}\vspace{5mm}}
\newcommand{\address}[1]{\vbox{\center\em#1}}
\newcommand{\email}[1]{\vbox{\center\tt#1}\vspace{5mm}}

\begin{document}

\singlespacing

\begin{titlepage}
\rightline{}
\bigskip
\bigskip\bigskip\bigskip\bigskip
\bigskip

\begin{center}

{\Large \bf {Collision in the interior of wormhole}}
\end{center}

\bigskip \noindent

\bigskip

\begin{center}

\author{Ying Zhao}

\address{Institute for Advanced Study,  Princeton, NJ 08540, USA }

\email{zhaoying@ias.edu}

\bigskip

\vspace{1cm}
\end{center}

\begin{abstract}

The Schwarzschild wormhole has been interpreted as an entangled state. If Alice and Bob fall into each of the black hole, they can meet in the interior. We interpret this meeting in terms of the quantum circuit that prepares the entangled state. Alice and Bob create growing perturbations in the circuit, and we argue that the overlap of these perturbations represents their meeting. We compare the gravity picture with circuit analysis, and identify the post-collision region as the region storing the gates that are not affected by any of the perturbations.

\medskip
\noindent
\end{abstract}

\end{titlepage}

\starttext \baselineskip=17.63pt \setcounter{footnote}{0}

{\hypersetup{hidelinks}
\tableofcontents
}

\section{Introduction}

 The Schwarzschild wormhole has been interpreted as an entangled state \cite{Maldacena:2001kr}. We assume Alice and Bob share the state, each holds one half of the entangled pairs. If each of them sends in a signal at appropriate time, the two signals can meet in the interior. On the other hand there are no boundary interactions between them. This makes ER = EPR \cite{Maldacena:2013xja} a mysterious statement.
 
 In AdS/CFT, it was argued that the bulk geometry reflects the quantum circuit preparing the boundary state \cite{Swingle:2009bg}\cite{Hartman:2013qma}\cite{Susskind:2014moa}. In particular, in the black hole interior there is a unitary circuit preparing the state whose complexity gives the volume / action in the interior \cite{Stanford:2014jda}\cite{Roberts:2014isa}\cite{Brown:2015bva}. 

\begin{figure}[H] 
 \begin{center}                      
      \includegraphics[width=4in]{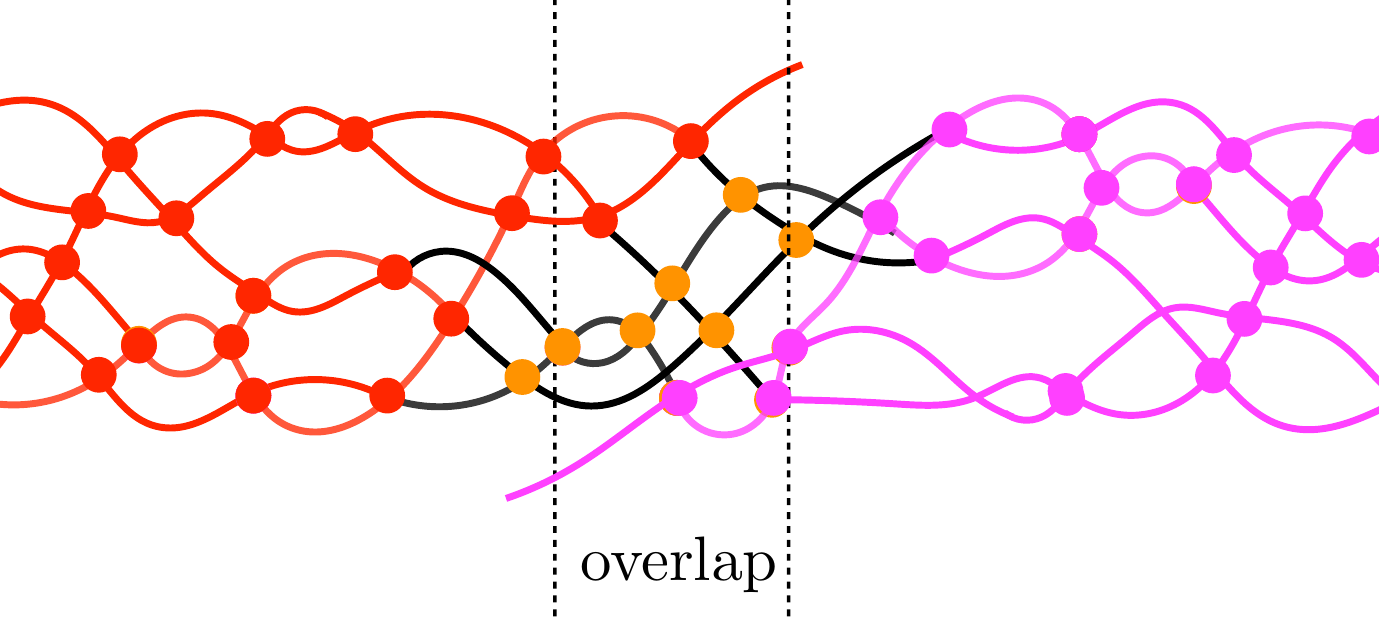}
      \caption{The two perturbations have overlap in the quantum circuit. }
  \label{circuit_50}
  \end{center}
\end{figure}

In this paper we use this quantum circuit picture to explain the ``meeting" of two signals from different boundaries. When Alice and Bob send in signals, they create growing perturbations in the circuit. When both signals are sent in early enough, the two perturbations will have overlap in the quantum circuit (Figure \ref{circuit_50}). Here, by ``overlap" we mean that there will be a portion of a circuit in which both perturbations appear. We argue that this overlap represents the meeting of the two signals in the interior.

In \cite{Zhao:2017isy} it was argued that different kinds of gates in the quantum circuit are stored in different parts of spacetime region. Say, Alice and Bob share thermofield double. Alice throws in a perturbation from left side at time $t_{w}$ (Figure \ref{Penrose_0}). We argued that from Alice' point of view, the interior geometry matches the quantum circuit in Figure \ref{geodesic}. In particular, the interior region inside Alice's entanglement wedge stores the quantum gates that are affected by the perturbation. 

In the first half of the paper we look at the same situation from Bob's point of view. We argue that we can also identify the trajectory of the infalling object in the interior close to the horizon as the perturbation in the quantum circuit (Figure \ref{Penrose_1_shock_2}). The closer the object is to the horizon, the larger its size is.

In the second half of the paper we consider the collision of two infalling objects in the interior of thermofield double, one sent in from the left boundary and one from the right boundary. In the corresponding quantum circuit, the two perturbations will have overlap (Figure \ref{circuit_50}). Using the epidemic model \cite{Hayden:2007cs}\cite{Susskind:2014jwa} with two epidemics coming in at different times, we estimate the number of the healthy gates in the overlap region, and show that it matches the spacetime volume of post-collision region in the gravity picture.

This paper is organized as follows. In section \ref{perturbed_thermofield_double} we study the example of perturbed theormofield double from the point of view of both boundaries. We identify the trajectory of infalling object close to the horizon in the interior as corresponding to the perturbation in the quantum circuit. In section \ref{collision} we study the collision of two infallling objects in the interior, one from the left boundary and the other one from the right boundary. We show that it corresponds to the overlap between the two perturbations in the quantum circuit. We find a detailed match between the volume of the post-collision region and the number of healthy gates from circuit analysis. In section \ref{discussion} we point out unanswered questions and future directions.

\section{Perturbed theomofield double and quantum circuit}
\label{perturbed_thermofield_double}

\subsection{Bulk tensor network and quantum circuit}

\begin{figure}[H] 
 \begin{center}                      
      \includegraphics[width=6in]{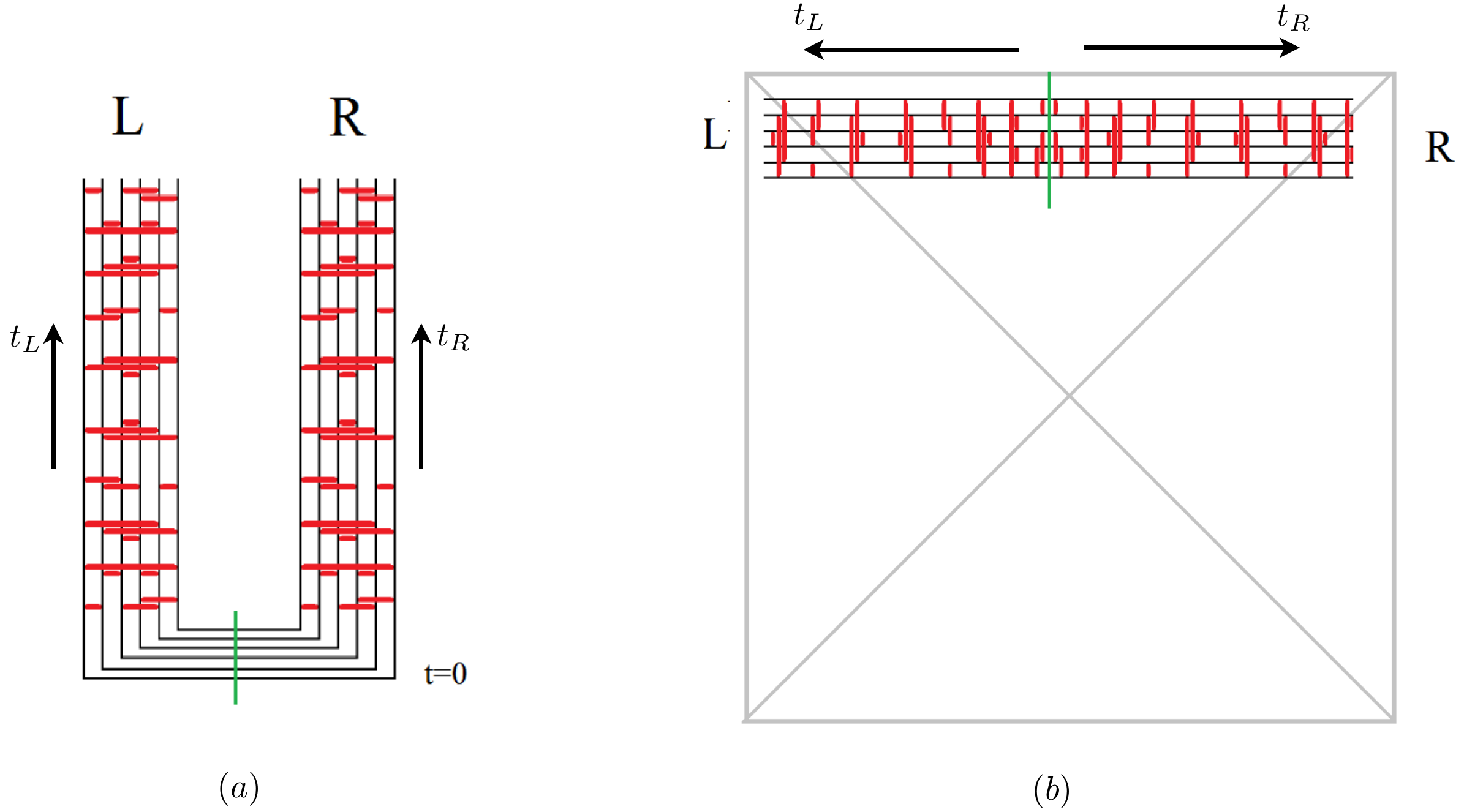}
      \caption{The quantum circuit in (a) becomes bulk tensor network in (b) in the interior.}
  \label{picture_1}
  \end{center}
\end{figure}

In this section we briefly review the translation between boundary quantum circuit and bulk geometry. Figure \ref{picture_1} is taken from \cite{Susskind:2014moa}\footnote{I thank Leonard Susskind for allowing me to use his figures.}.
 
Susskind pointed out that once we lift the quantum circuit in Figure \ref{picture_1}(a) and lay it on a spatial slice in Figure \ref{picture_1}(b), we get a tensor network description of wormhole \cite{Susskind:2014moa}. The gates in the quantum circuit of Figure \ref{picture_1} become tensors in the tensor network representing the wormhole (ERB). 

We will use the convention that the circuit time $\tau$ increases toward the right, so we will identify the circuit time with right time $t_R$, and minus the left time $-t_L$.

\subsection{Epidemic model}

\begin{figure}[H] 
 \begin{center}                      
      \includegraphics[width=3in]{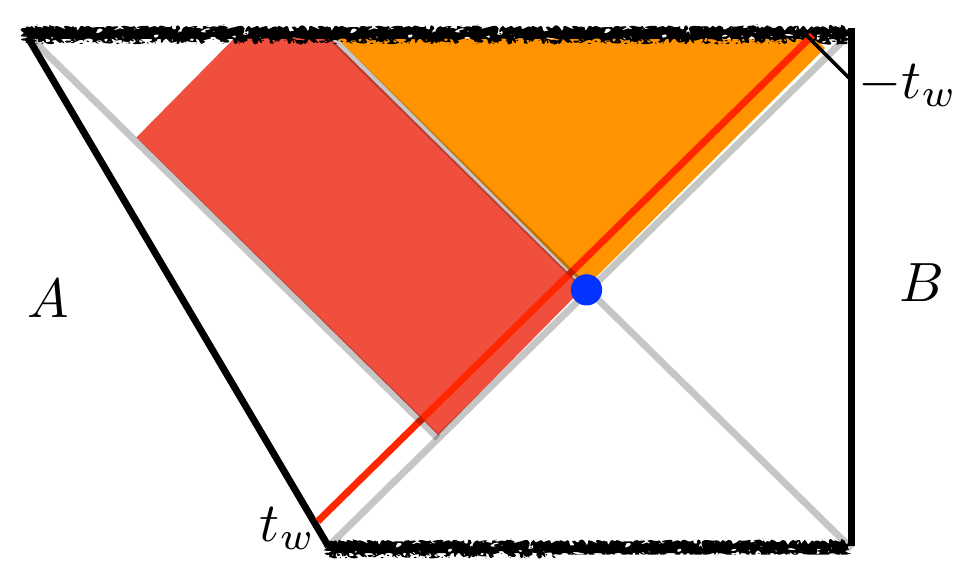}
      \caption{Thermofield double perturbed by Alice at time $t_w$}
  \label{Penrose_0}
  \end{center}
\end{figure}

Let's start from a situation where Alice and Bob share the two sides of thermofield double. Alice has the left side while Bob has the right side. Alice throws in a perturbation from the left side (Figure \ref{Penrose_0}). We want to compare the geometry and quantum circuit from both the point of view of Alice and Bob. 

In Figure \ref{Penrose_0}, Alice sent in a qubit from her boundary at time $t_{wL}$. As a result the left horizon expands. The red region in Figure \ref{Penrose_0} is the region between the new horizon after the perturbation and the old horizon without the perturbation. The RT surface stays at the blue dot.

The epidemic model was introduced to describe the perturbation of a black hole \cite{Susskind:2014jwa}\cite{Roberts:2014isa}. We briefly review it here as it will play a crucial role in the rest of our discussion. We model the black hole by $S$ qubits and we model the dynamics by a Hayden-Preskill type circuit \cite{Hayden:2007cs}: At each time step the qubits are randomly grouped into $\frac{S}{2}$ pairs, and on each pair a randomly chosen $2$-qubit gate is applied. We can characterize the effect of some small perturbation in such a system as follows. Imagine the unpertubed system contains $S$ healthy qubits, and the perturbation is one extra qubit carrying some disease. The sick qubit enters the system at $\tau = 0$. Any qubits who interact directly or indirectly with sick qubits will get sick. We define the size of the epidemic $s_{ep}(\tau)$ to be the number of sick qubits at time $\tau$ where $\tau$ is discrete circuit time.\footnote{Later we will identity the circuit time and Schwarzshild time: $d\tau = \frac{2\pi}{\beta}dt$. } It satisfies
\begin{align}
&\frac{ds_{ep}(\tau)}{d\tau} = \frac{(S+1-s_{ep})s_{ep}}{S}\nonumber\\
&\frac{s_{ep}(\tau)}{S+1} = \frac{\frac{\delta S}{S}e^{\frac{S+1}{S}\tau}}{1+\frac{\delta S}{S}e^{\frac{S+1}{S}\tau}} = \frac{\frac{\delta S}{S}e^{\tau}}{1+\frac{\delta S}{S}e^{\tau}}\label{size_epidemic}
\end{align}
We used initial condition $s_{ep}(\tau = 0) = \delta S$ where $\delta S$ is the number of initially infected qubits. It is proportional to the increase of thermal entropy from the perturbation. Later we will adjust the prefactor to make a better match with the gravity picture. 

Assume Alice and Bob share thermofield double and Alice applies the perturbation. At each time step from $\tau-\Delta\tau$ to $\tau$, $S$ gates are applied on Alice's side.\footnote{We count each 2-qubit gate as $2$ gates.} Among them, $s_{ep}(\tau)$ gates involve the extra qubit and cannot be undone by Bob. The remaining $S-s_{ep}(\tau)$ gates can be undone by Alice as well as Bob. We use $N_{sick}$ to denote the total number gates affected by the extra qubit in the circuit, i.e., the gates that can only by undone by Alice but not by Bob. We use $N_{healthy}$ to denote the total number of unaffected the gates, i.e., the gates that can be undone by Alice as well as Bob. We have
\begin{align*}
	\frac{d N_{sick}}{d\tau} = s_{ep}(\tau),\ \ \ \frac{dN_{healthy}}{d\tau} = S-s_{ep}(\tau)
\end{align*}

\subsection{Quantum circuit from the point of view of Alice}

\begin{figure}[H] 
 \begin{center}                      
      \includegraphics[width=5in]{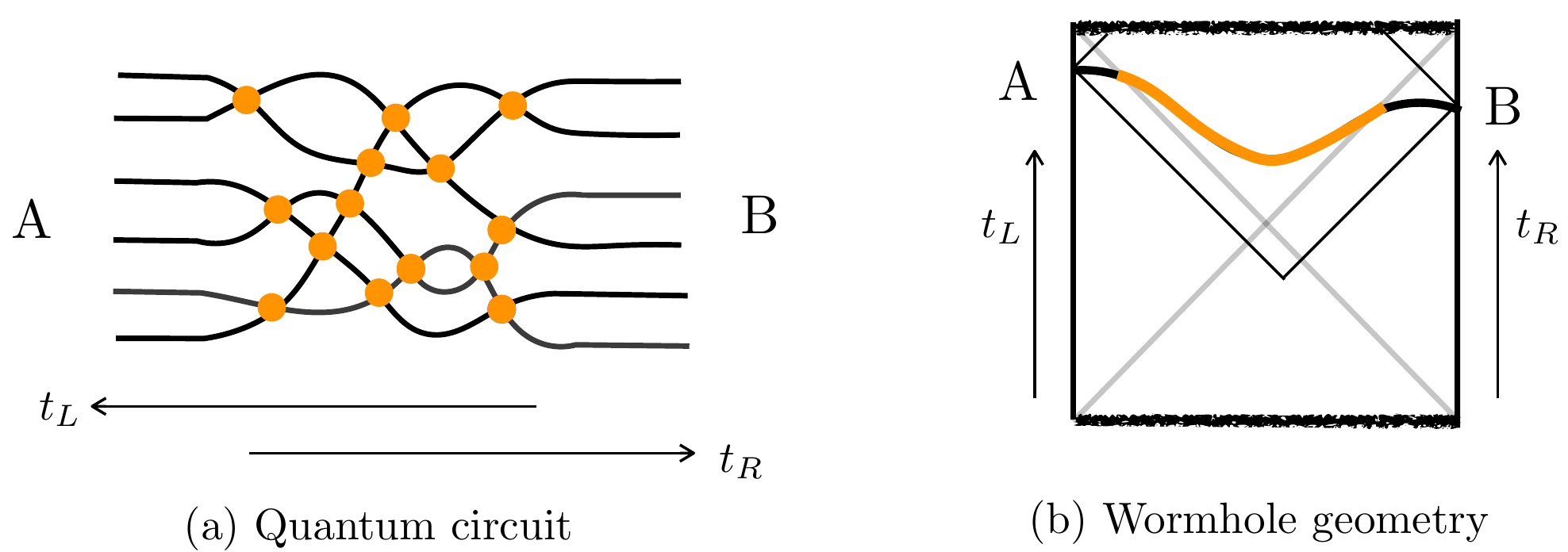}
      \caption{The wormhole grows as the circuit grows.}
  \label{thermofield_double_time_evolved}
  \end{center}
\end{figure}

We first look at the quantum circuit without perturbations. We represent the black hole by $S$ Bell pairs. As the left (right) boundary time increases, the circuit grows toward left (right) (Figure \ref{thermofield_double_time_evolved}(a)). The wormhole also grows (Figure \ref{thermofield_double_time_evolved}(b)). 

Now Alice sends in some perturbation from left boundary at time $t_w$. In the quantum circuit picutre, we model Alice' perturbation by throwing in an extra qubit. Here is the reason: Alice' perturbation cannot be undone by Bob, so it changes Alice' density matrix. In particular, Alice' density matrix will no longer be exactly thermal, and its thermal entropy will be larger than fine-grained entropy. We use this extra qubit to represent the increase of thermal entropy.

\begin{figure}[H] 
 \begin{center}                      
      \includegraphics[width=5in]{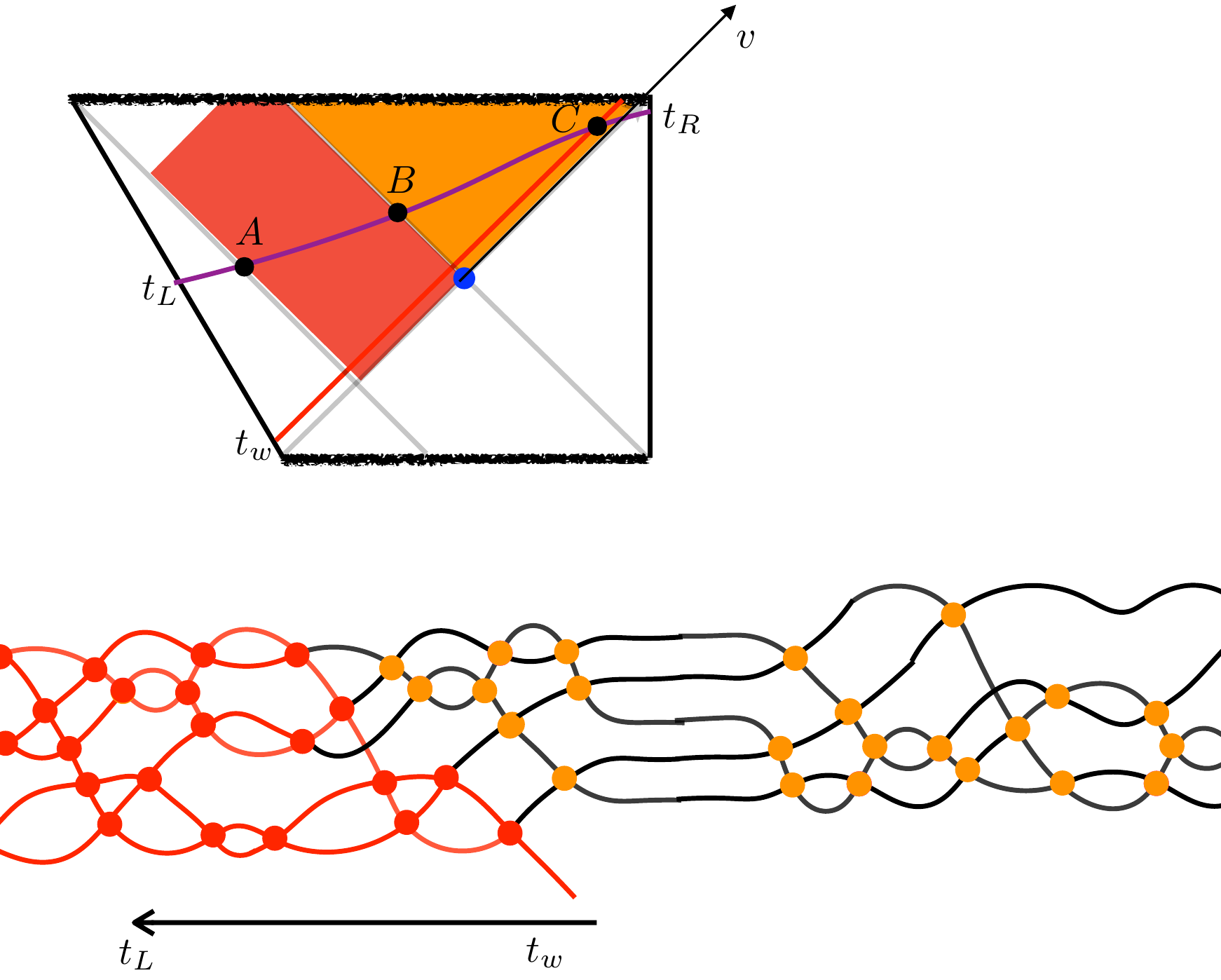}
      \caption{Penrose diagram and quantum circuit of perturbed thermofield double}
  \label{geodesic}
  \end{center}
\end{figure}

In \cite{Zhao:2017isy} it was argued that the gates in the quantum circuit that can be undone by both Alice an Bob are stored in the shared entanglement region (orange region in Figure \ref{geodesic}), while the gates that can be undone by Alice but not by Bob are stored in Alice's entanglement wedge (red region in Figure \ref{geodesic}). The spacetime volumes in these regions were compared with the epidemic model of quantum circuit. The exact quantity to look at doesn't matter. One can instead look at the geodesic distance between the two boundaries.

We fix some very large right time $t_{R}\rightarrow \infty$ and vary the left time $t_L$. Instead of looking at the entire geodesic between the two boundaries, we focus on the part of the geodesic contained in the orange region as well as the red region in Figure \ref{geodesic}. We want to say that the growth of the orange region (geodesic BC) corresponds to the growth of the orange gates in the circuit in Figure \ref{geodesic} (the gates that can be undone by both Alice and Bob), while the growth of the red region (geodesic AB) corresponds to the growth of the red gates (the gates that can be undone by Alice but not by Bob). In BTZ black hole geometry, one can show that (appendix \ref{geodesic_cal_1})
\begin{align*}
&\frac{d}{dt_L}\left(\frac{d_{AB}}{l}\right) =\frac{2\pi}{\beta} \frac{e^{\frac{2\pi}{\beta}(t_L-t_w-t_*)}}{1+e^{\frac{2\pi}{\beta}(t_L-t_w-t_*)}} = \frac{2\pi}{\beta}\qty(\frac{s_{ep}[\frac{2\pi}{\beta}(t_L-t_w)]}{S}) = \frac{d}{dt_L}\qty(\frac{N_{sick}}{S})\\
&\frac{d}{dt_L}\left(\frac{d_{BC}}{l}\right) = \frac{2\pi}{\beta}\frac{1}{1+e^{\frac{2\pi}{\beta}(t_L-t_w-t_*)}}=\frac{2\pi}{\beta}\qty(1-\frac{s_{ep}[\frac{2\pi}{\beta}(t_L-t_w)]}{S}) = \frac{d}{dt_L}\qty(\frac{N_{healthy}}{S})
\end{align*}
where $s_{ep}(\tau)$ is the size function in epidemic model as in \eqref{size_epidemic}.
The above geodesic length calculation shows perfect match with the epidemic model. The fact that the match is perfect may be an artifact of the special geometry of BTZ as well as its geodesic distance. The lesson is that the growth of certain spacetime regions shows the same features of time dependence as the growth of certain kinds of gates in the quantum circuit, like exponential growth, saturation after scrambling time, etc.

\subsection{Quantum circuit from the point of view of Bob}

\begin{figure}[H] 
 \begin{center}                      
      \includegraphics[width=2.6in]{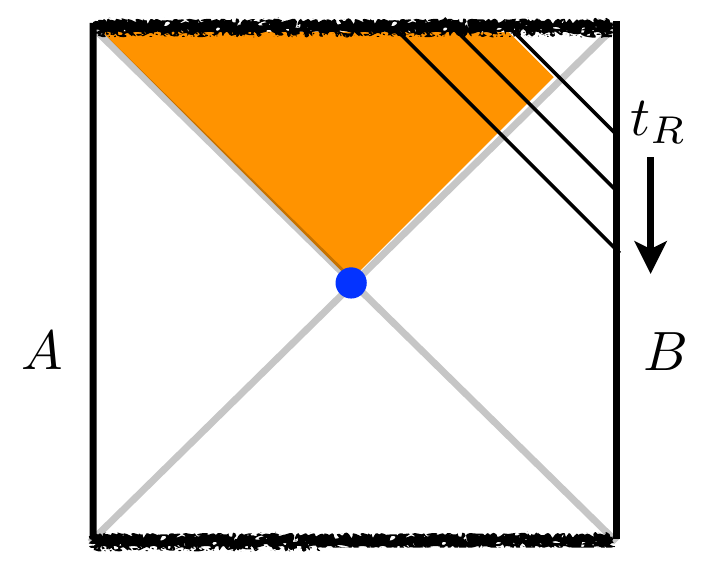}
      \caption{As Bob decreases the right time, he will undo the gates stored in the future interior (orange region) at a steady rate. }
  \label{Penrose_00}
  \end{center}
\end{figure}

Let's first consider the situation when there is no shockwave and Alice and Bob share standard thermofield double. We put left time at $t_L = +\infty$. As Bob decreases  the right time $t_R$, he will undo the gates stored in the future interior at a steady rate (Figure \ref{Penrose_00}). 

\begin{figure}[H] 
 \begin{center}                      
      \includegraphics[width=3in]{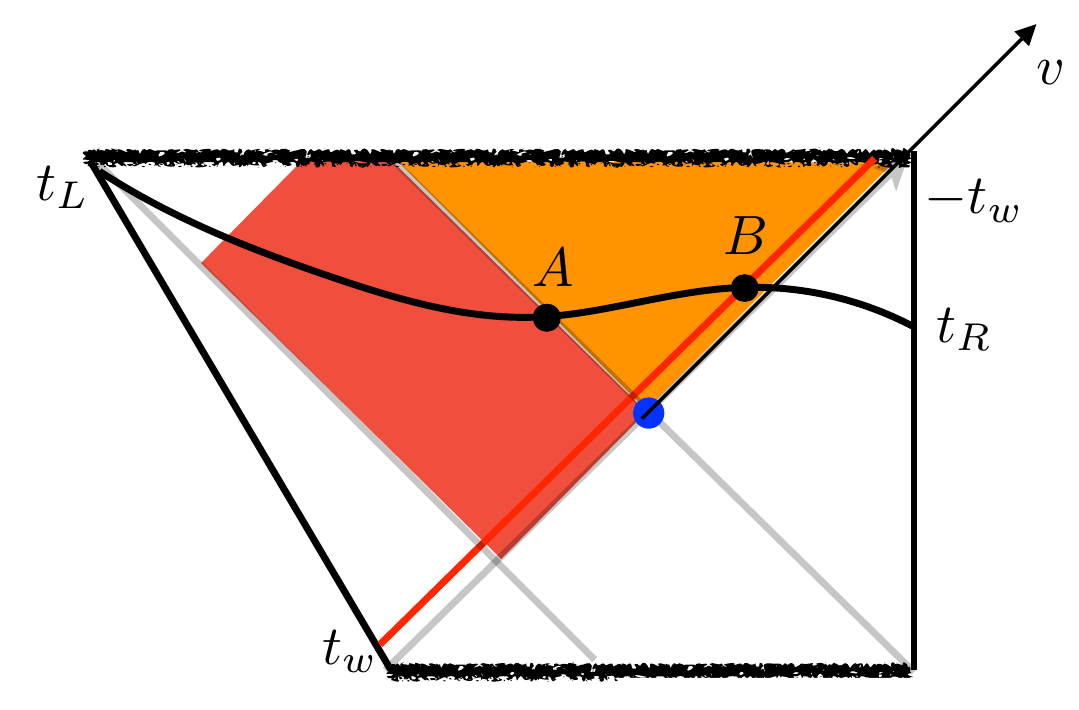}
      \caption{We fix the left time at future infinity and vary the right time.}
  \label{geodesic_2}
  \end{center}
\end{figure}

Now we consider the situation where Alice sent in a perturbation from her side. We look at the quantum circuit in Figure \ref{geodesic} from the point of view of Bob. What do we expect? Bob should be able to undo the orange gates but Bob cannot undo the red gates. The existence of those red gates will affect the rate at which Bob can undo the orange gates. We again look at the time dependence of geodesic inside the future entanglement region (orange region), but this time we fix the left time at $+\infty$ and vary the right time (Figure \ref{geodesic_2}). Here is its time dependence we found in BTZ blackground (appendix \ref{geodesic_cal_2}).

\begin{align}
	\frac{d_{\text{AB}}}{l} =\ & \frac{2\pi}{\beta}(t_R+t_w+t_*)+\log(1+\alpha e^{-\frac{2\pi}{\beta}t_R})\nonumber\\
	\frac{d}{dt_R}\frac{d_{\text{AB}}}{l} =\ & \frac{2\pi}{\beta}\qty[1-\frac{e^{\frac{2\pi}{\beta}(-t_w-t_*-t_R)}}{1+e^{\frac{2\pi}{\beta}(-t_w-t_*-t_R)}}]  = \frac{2\pi}{\beta}\qty(1-\frac{s_{ep}[\frac{2\pi}{\beta}(-t_{w}-t_R)]}{S})\label{geodesic_Bob}
\end{align}

\begin{figure}[H] 
 \begin{center}                      
      \includegraphics[width=3in]{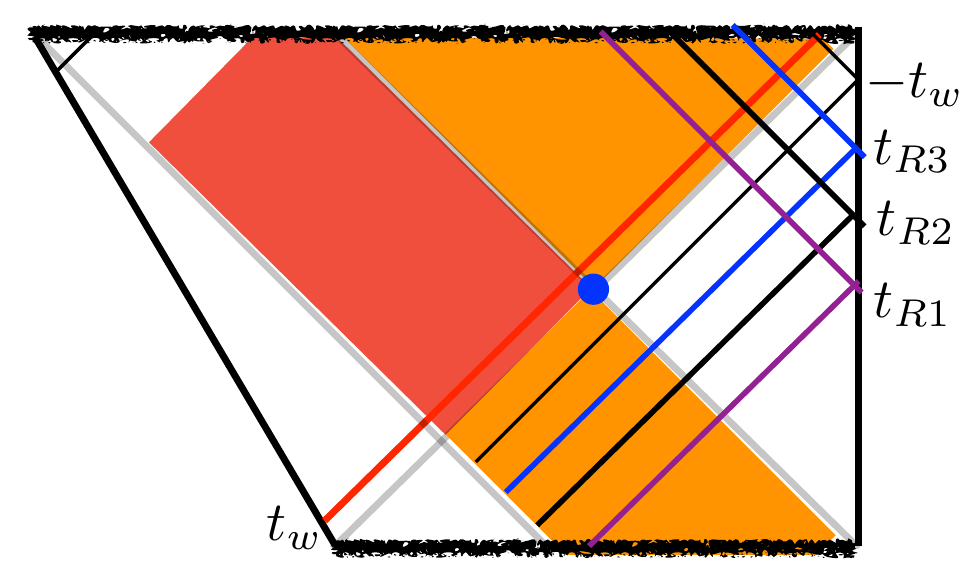}
      \caption{During each time step as we decrease the right time, $S-s_{ep}$ gates are cancelled by the gates in the future interior. The remaining $s_{ep}$ gates are stored in the past interior.}
  \label{WDW}
  \end{center}
\end{figure}

So far we only looked at the time dependence of the length of geodesic in the future interior region. If we take the point of view that the bulk dual to a boundary state is the entire Wheeler-DeWitt (WDW) patch, we can say more. Imagine that during the time $t_R$ and $t_R-\Delta t$, $S\Delta t$ gates are applied in the right CFT. Among them, $S\Delta t(1-\frac{s_{ep}}{S})$ gates are cancelled by the gates stored in the future interior. Then what happens to the remaining $s_{ep}\Delta t$ gates? They are stored in the past interior. See Figure \ref{WDW} \footnote{I thank Juan Maldacena for correcting me on this.}.

\begin{figure}[H] 
 \begin{center}                      
      \includegraphics[width=6in]{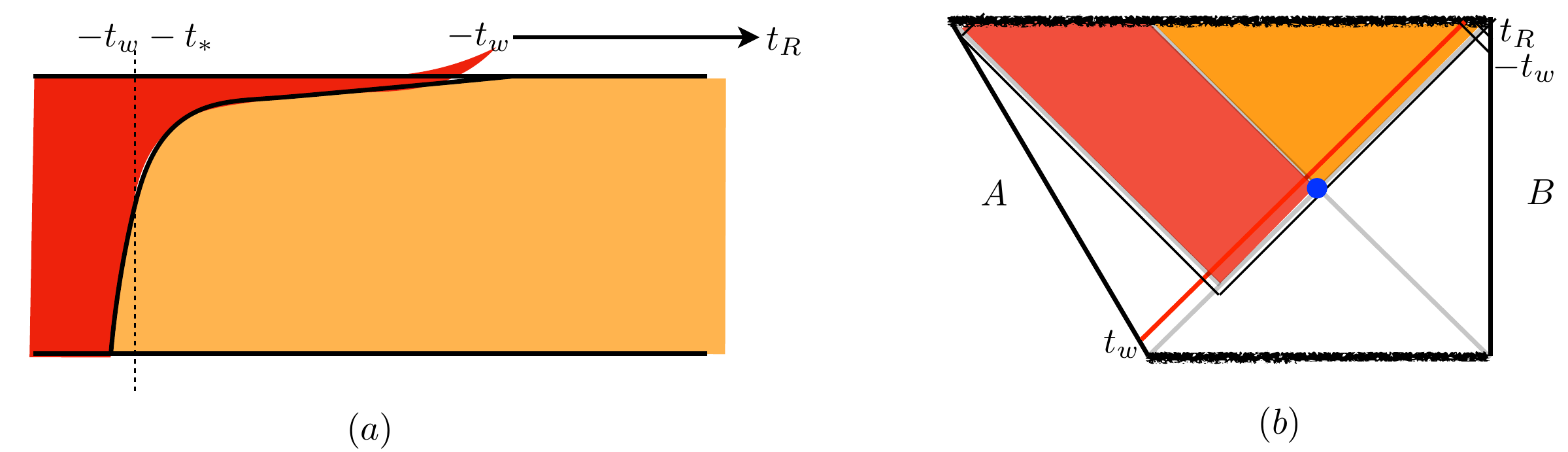}
      \caption{Quantum circuit and WDW patch at $t_R>-t_w$.}
  \label{circuit_fold_0}
  \end{center}
\end{figure}

Here are comparisons of the quantum circuit and WDW patch at different times. With fixed $t_L$ large, Figure \ref{circuit_fold_0} shows the quantum circuit and WDW patch in Penrose diagram at $t_R>-t_w$. 

\begin{figure}[H] 
 \begin{center}                      
      \includegraphics[width=5.3in]{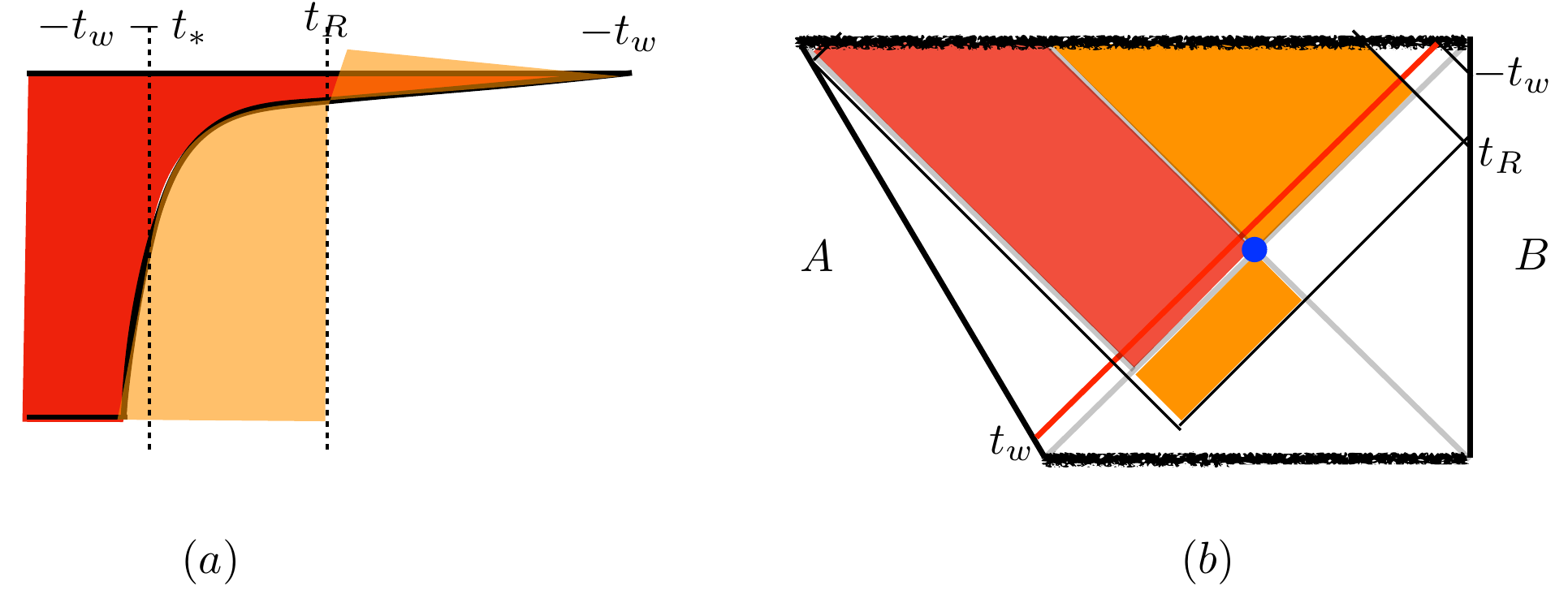}
      \caption{Quantum circuit and WDW patch at $-t_w-t_*<t_R<-t_w$.}
  \label{circuit_fold_1}
  \end{center}
\end{figure}

Now Bob decrease the right time $t_R$. He can undo the orange gates in Figure \ref{circuit_fold_0}(a), but he cannot undo the red gates. Instead, he will create a fold on top of the red gates (Figure \ref{circuit_fold_1}(a)).\footnote{This was pointed out to me by Juan Maldacena.} This fold is stored in the past interior (Figure \ref{circuit_fold_1}(b)).

\begin{figure}[H] 
 \begin{center}                      
      \includegraphics[width=5.3in]{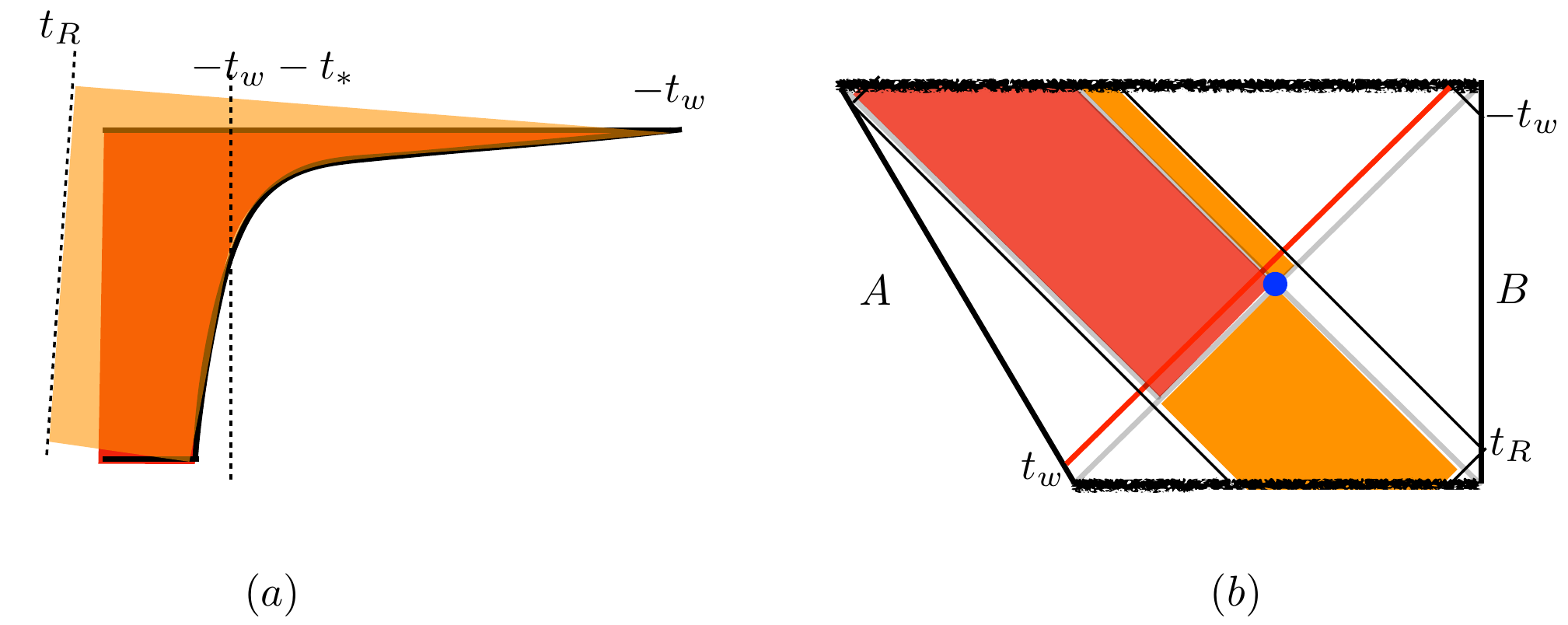}
      \caption{Quantum circuit and WDW patch at $t_R<-t_w-t_*$.}
  \label{circuit_fold_2}
  \end{center}
\end{figure}

If Bob further decreases the right time until $t_R<-t_w-t_*$, all the orange gates in Figure \ref{circuit_fold_0}(a) have been undone. The gates in the extra fold are stored in the past interior (Figure \ref{circuit_fold_2}).  

\subsection{Size and interior trajectory}

We see that since Bob cannot undo the red gates in the quantum circuit in Figure \ref{geodesic}, the complexity has non-trivial time dependence as he changes the right time. 

\begin{figure}[H] 
 \begin{center}                      
      \includegraphics[width=4in]{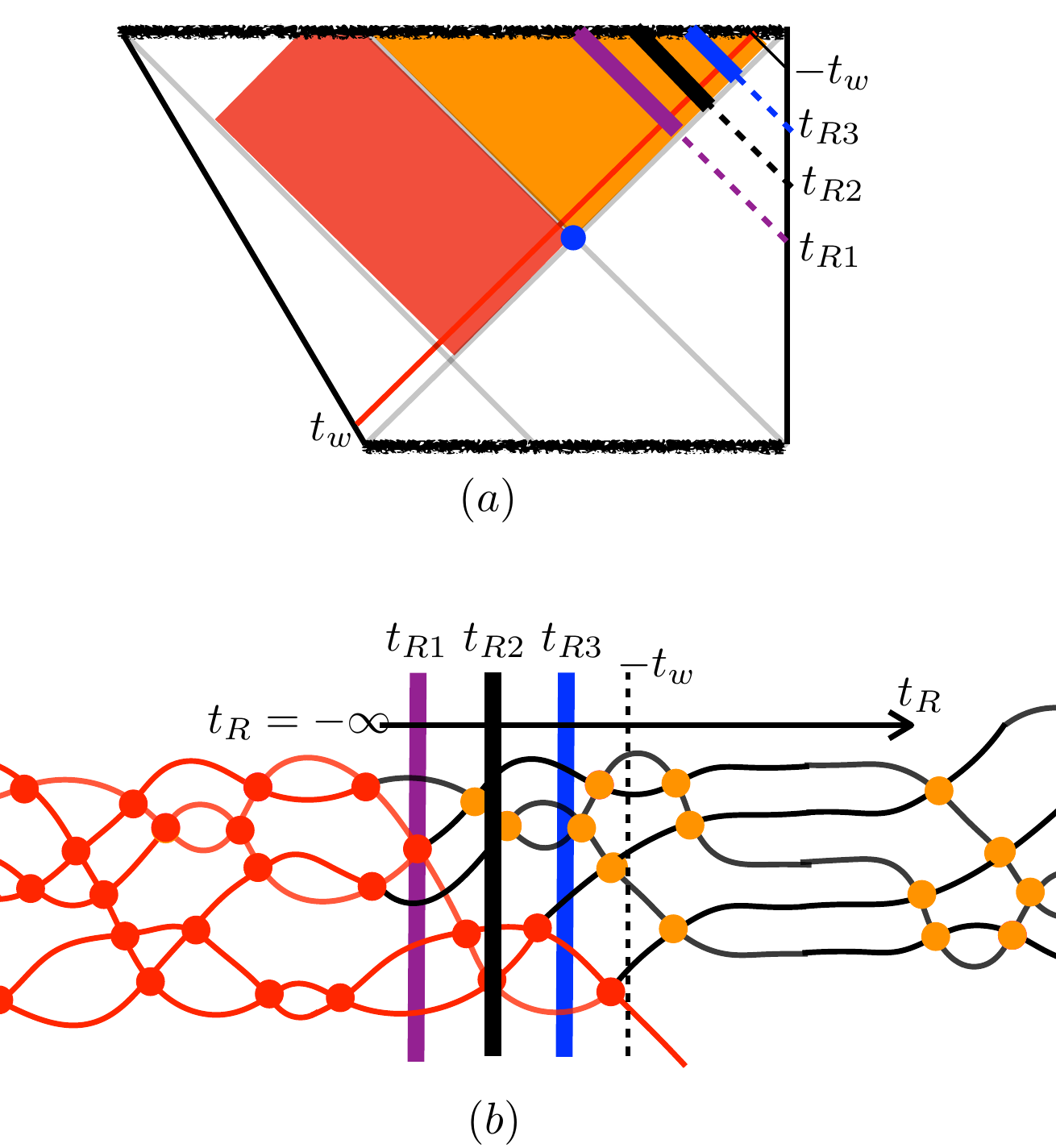}
      \caption{When Bob decreases the right time, he is scanning through the quantum circuit from the right to the left. }
  \label{Penrose_1_shock}
  \end{center}
\end{figure}

\begin{figure}[H] 
 \begin{center}                      
      \includegraphics[width=3in]{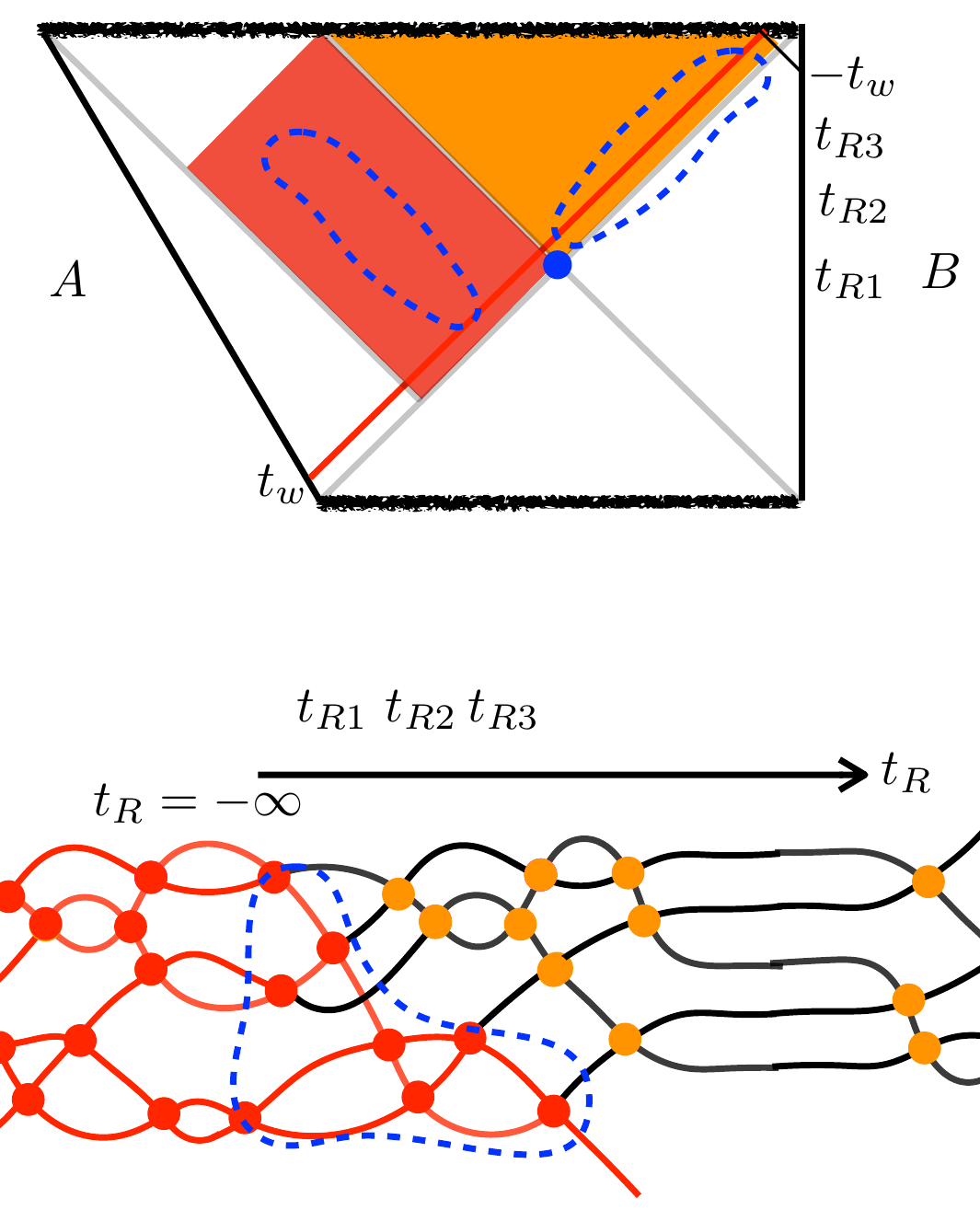}
      \caption{We can identify the part of the interior shockwave trajectory near the horizon as the perturbation in the quantum circuit.}
  \label{Penrose_1_shock_2}
  \end{center}
\end{figure}

The time dependence of the geodesic length in \eqref{geodesic_Bob} shows that when Bob decreases the right time, he is scanning through the quantum circuit from the right to the left (Figure \ref{Penrose_1_shock}). He will encounter the perturbation at $t_R = -t_w$. 
In Figure \ref{Penrose_1_shock}(a) we draw three sections in the Penrose diagram, which correspond to three sections in the quantum circuit of the same color in Figure \ref{Penrose_1_shock}(b). In this sense one can identify the part of the shockwave trajectory in the interior near the horizon as the perturbation in the quantum circuit (Figure \ref{Penrose_1_shock_2}). The closer the object is to the horizon, the larger its size is. 

If we look at the Penrose diagram in Figure \ref{Penrose_1_shock_2}, we see that both the red region inside Alice's entanglement wedge and the trajectory of the infalling object inside the orange region have to do with the perturbation in the quantum circuit. We circled them out in blue dashed circles. We see complementarity at play here \cite{Susskind:1993if}. More detailed discussion about this is the topic of current investigation \cite{Haehl:2021toappear}.

\section{Collision in the interior}
\label{collision}

\begin{figure}[H] 
 \begin{center}                      
      \includegraphics[width=3in]{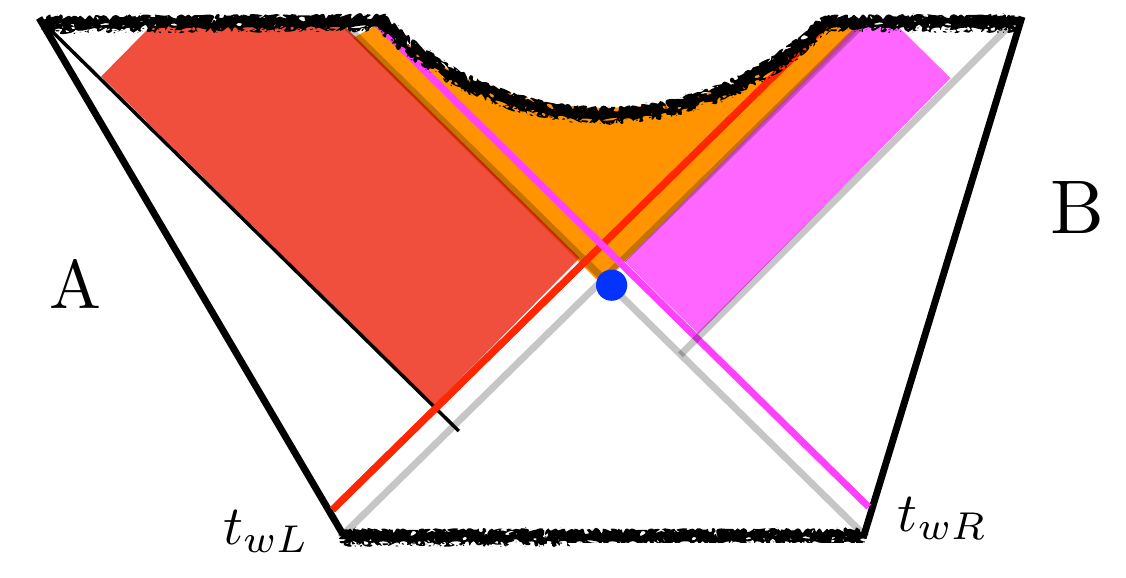}
      \caption{Infalling obejcts from the two boundaries can meet in the interior.}
  \label{Penrose_2}
  \end{center}
\end{figure}

It's a mysterious feature in ER = EPR that infallling objects from the two boundaries can meet in the shared interior (Figure \ref{Penrose_2}), as there are no interactions between them in the boundary theory. Here, we give a quantum circuit interpretation of this phenomenon. 

\subsection{The spreading of two epidemics: overlap of two perturbations in the quantum circuit}

\begin{figure}[H] 
 \begin{center}                      
      \includegraphics[width=4.6in]{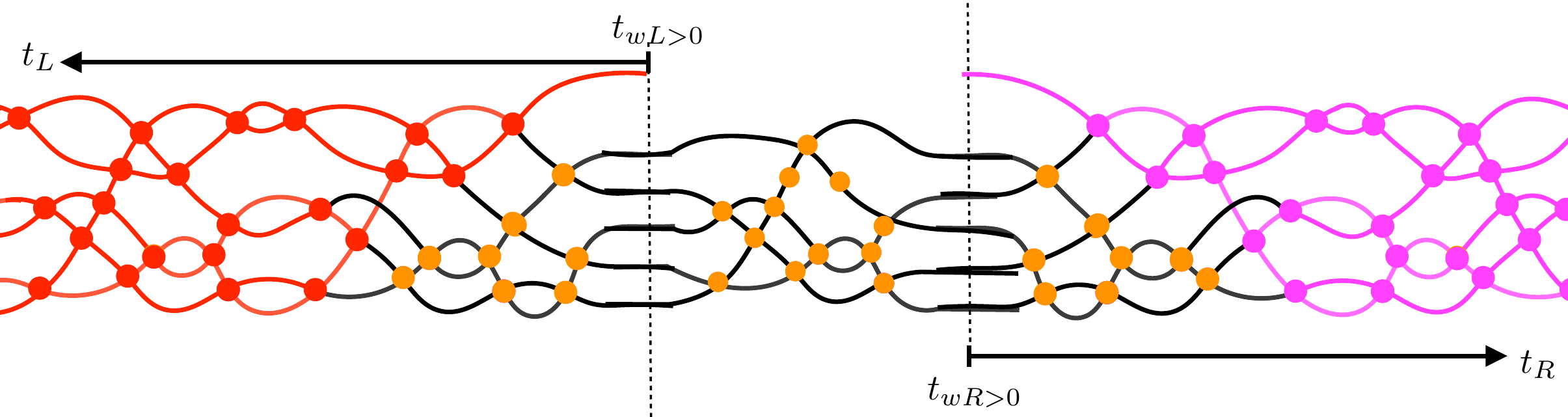}
      \caption{The two perturbations do not overlap in the quantum circuit. }
  \label{circuit_40}
  \end{center}
\end{figure}

Let's start from thermofield double shared between Alice and Bob. We model the quantum circuit by $S$ maximally entangled Bell pairs. Now Alice throws in a qubit from the left boundary at time $t_{wL}$, and Bob throws in a qubit from the right boundary at time $t_{wR}$. We first assume $t_{wL}>0$ and $t_{wR}>0$. In the corresponding quantum circuit, there will be no overlaps between the two perturbations (Figure \ref{circuit_40}). 

\begin{figure}[H] 
 \begin{center}                      
      \includegraphics[width=2.6in]{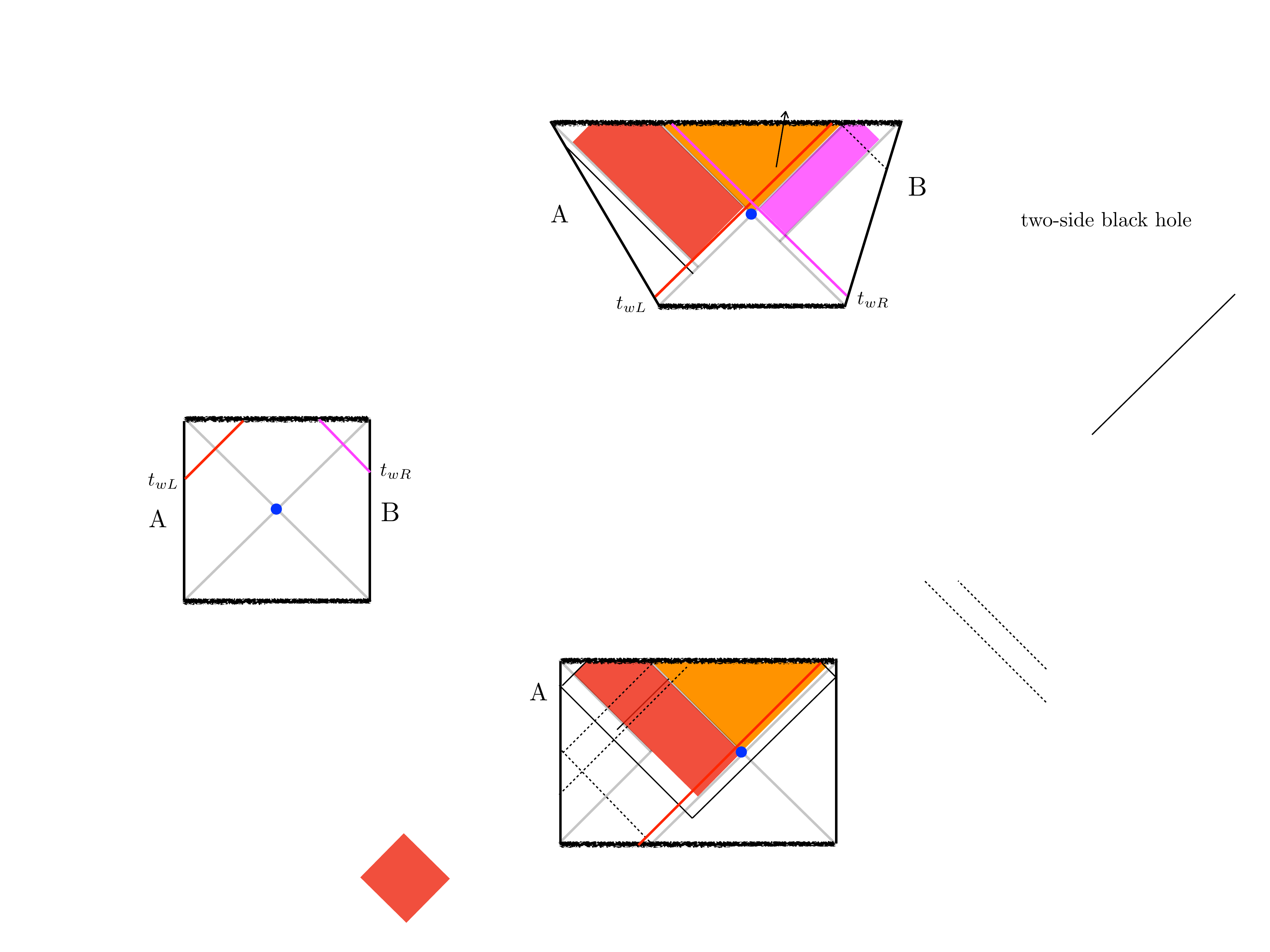}
      \caption{The two infalling objects do not meet in the interior. }
  \label{Penrose_2shock_1}
  \end{center}
\end{figure}

Correspondingly, the two perturbations do not collide in the interior (Figure \ref{Penrose_2shock_1}). 

\begin{figure}[H] 
 \begin{center}                      
      \includegraphics[width=4in]{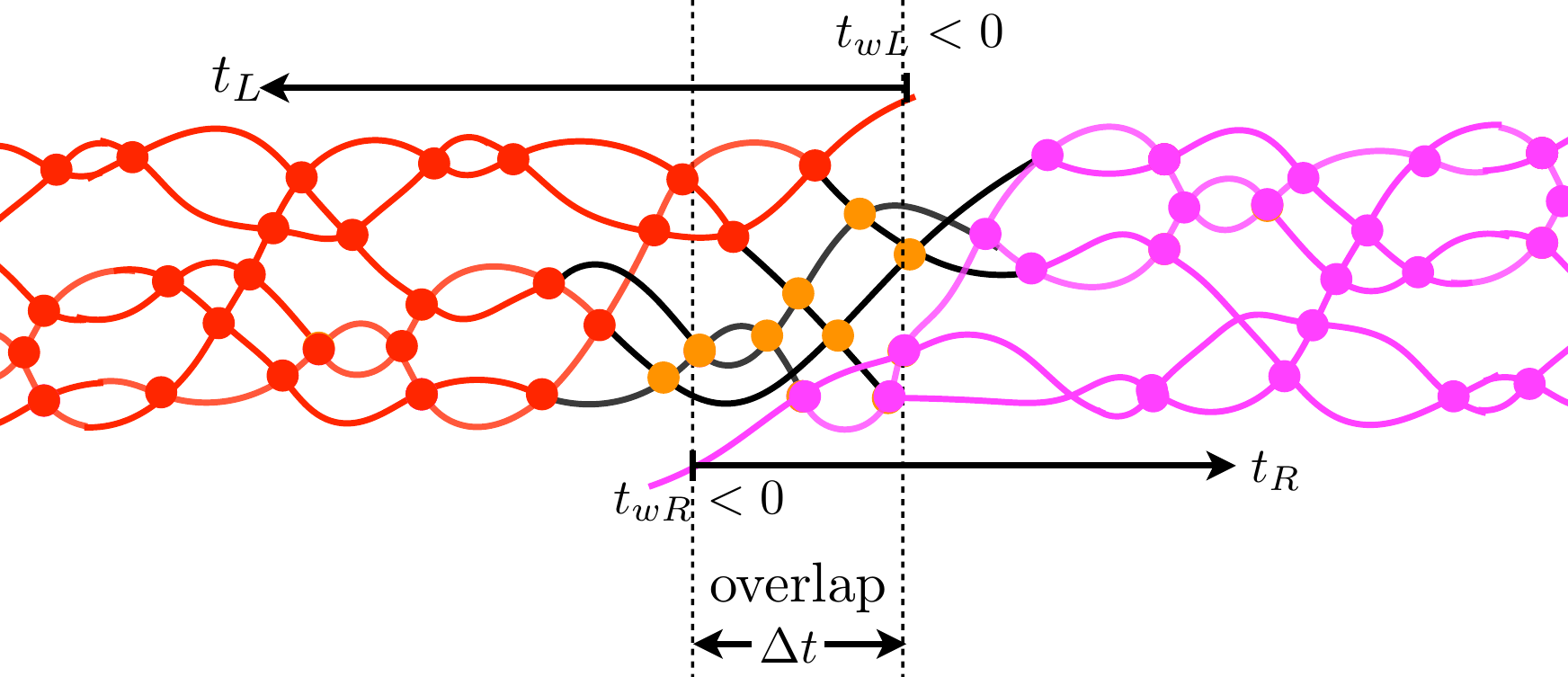}
      \caption{The two perturbations have overlap in the quantum circuit. }
  \label{circuit_5}
  \end{center}
\end{figure}

Next we assume $t_{wL}<0$ and $t_{wR}<0$. The two perturbations will have overlap in the quantum circuit (Figure \ref{circuit_5}). 

Correspondingly, the two perturbations will collide in the interior (Figure \ref{Penrose_2}). \\

Imagine we have a quantum circuit made of orange gates that can be undone from both sides. We call these orange gates healthy gates. Now a red epidemic comes in at time $-t_{wL}>0$ \footnote{Recall that we identify the circuit time with the right time $t_R$, and with minus the left time $-t_L$. } and propagates toward the left of the circuit. Also, a purple epidemic comes in at time $t_{wR}<0$ and propagates toward the right (Figure \ref{circuit_5}). In the next section, we will estimate the number of healthy gates in the overlaping region of the circuit and compare it with the volume of the post-collision region in gravity picture.

\subsection{Post-collision region and the number of healthy gates in the circuit}

We look at the quantum circuit model with two overlaping epidemics. Let $\Delta t\equiv -t_{wL}-t_{wR}$ be the overlapping time of the two epidemics (Figure \ref{circuit_5}). 

We estimate the number of healthy gates in the overlap region of the quantum circuit. In Figure \ref{circuit_overlap_1}, the probability of being healthy from the red epidemic is $(1-\frac{s_{ep}^{red}}{S}) = \frac{1}{1+\frac{\delta S_1}{S}e^{\frac{2\pi}{\beta}(-t_{wL}-t)}}$. The probability of being healthy from the purple epidemic is $(1-\frac{s_{ep}^{purple}}{S}) = \frac{1}{1+\frac{\delta S_2}{S}e^{\frac{2\pi}{\beta}(t-t_{wR})}}$. The probability of being healthy from both epidemics in the overlap region is the product of these two.

\begin{enumerate}

\item{Early time: $\frac{\beta}{2\pi}\ll \Delta t \ll t_*$}

\begin{figure}[H] 
 \begin{center}                      
      \includegraphics[width=3in]{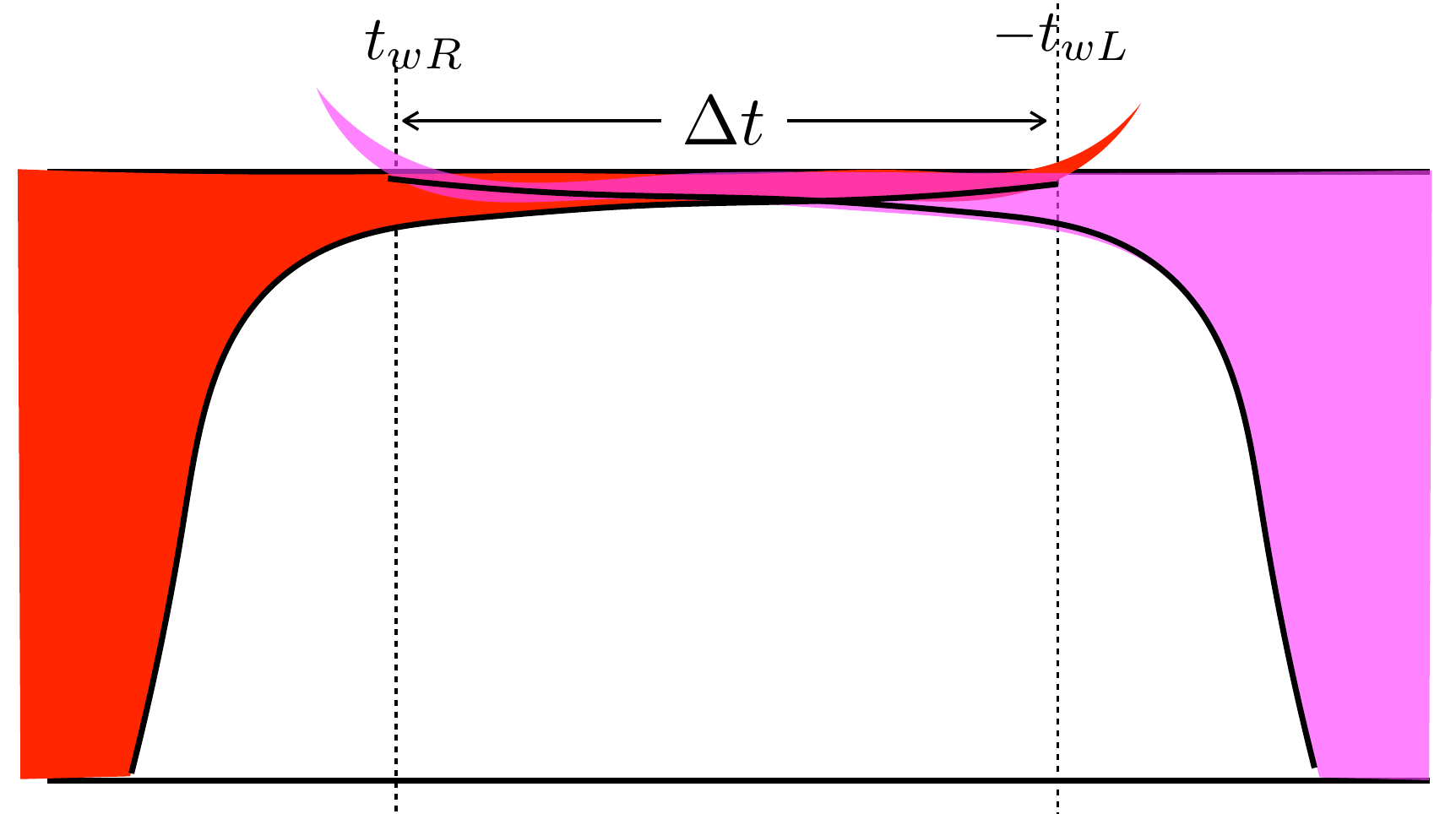}
      \caption{$\frac{\beta}{2\pi}\ll\Delta t\ll t_*$. The red region represents the gates contaminated by the red epidemic. The purple region represents the gates contaminated by the purple epidemic. The white region represents the healthy gates.}
  \label{circuit_overlap_1}
  \end{center}
\end{figure}

The number of healthy gates in the overlap region is given by
\begin{align}
	\frac{N_{\text{healthy}}}{S} = \ &\int_{t_{wR}}^{-t_{wL}}\frac{2\pi}{\beta}dt\qty(\frac{1}{1+\frac{\delta S_1}{S}e^{\frac{2\pi}{\beta}(-t_{wL}-t)}})\qty(\frac{1}{1+\frac{\delta S_2}{S}e^{\frac{2\pi}{\beta}(t-t_{wR})}})\nonumber\\
	=\ &\frac{2\pi}{\beta}\Delta t-\frac{\delta S_1+\delta S_2}{S}e^{\frac{2\pi}{\beta}\Delta t}+\mathcal{O}\qty(\qty(\frac{1}{S}e^{\frac{2\pi}{\beta}\Delta t})^2)\label{circuit_early}
\end{align}

Now we look at the Penrose diagram. The collision happens at radius $r_c = r_h \tanh(\frac{\pi}{\beta}\Delta t)$. In the post collision region, a larger black hole forms. In BTZ geometry, the new horizon radius $\tilde r_h$ is given by \cite{Dray:1985yt}\cite{Shenker:2013yza}\footnote{This result depends on spherical symmetry as well as thin wall approximation.}
\begin{align*}
	\frac{\tilde r_h^2}{r_h^2} =1+\frac{2\delta S_1}{S}+\frac{2\delta S_2}{S}+\frac{4\delta S_1\delta S_2}{S^2}\cosh^2\qty(\frac{\pi}{\beta}\Delta t) 
\end{align*}

The spacetime volume in the post collision region is given by
\begin{align*}
	V  =\ &2\pi \tilde r_h l^2\qty(\frac{1}{2}\log\frac{\tilde r_h+r_c}{\tilde r_h-r_c}-\frac{r_c}{\tilde r_h})
\end{align*}

In the early regime where $\frac{\beta}{2\pi}\ll\Delta t\ll t_*$, we have
\begin{align}
	\frac{V}{\pi\tilde r_h l^2} =\ & \log\frac{1+\frac{\delta S_1}{S}+\frac{\delta S_2}{S}+\tanh(\frac{\pi}{\beta}\Delta t)}{1+\frac{\delta S_1}{S}+\frac{\delta S_2}{S}-\tanh(\frac{\pi}{\beta}\Delta t)}-2\frac{\tanh(\frac{\pi}{\beta}\Delta t)}{1+\frac{\delta S_1+\delta S_2}{S}}\nonumber\\
	\approx\ &\frac{2\pi}{\beta}\Delta t-\frac{\delta S_1+\delta S_2}{S}\sinh(\frac{2\pi}{\beta}\Delta t)\nonumber\\
	\frac{V}{\pi r_hl^2}\approx\ &\frac{2\pi}{\beta}\Delta t-\frac{\delta S_1+\delta S_2}{2S}e^{\frac{2\pi}{\beta}\Delta t}\label{volume_earlytime}
\end{align}

\eqref{circuit_early} and \eqref{volume_earlytime} agree up to a shift of $\Delta t$.

In this case, we can also look at the total number of healthy gates (the entire white region in Figure \ref{circuit_overlap_1}) and compare it with the volume of the entire future interior region shared between the two sides (orange region in Figure \ref{Penrose_2shock_3}). 
\begin{figure}[H] 
 \begin{center}                      
      \includegraphics[width=3in]{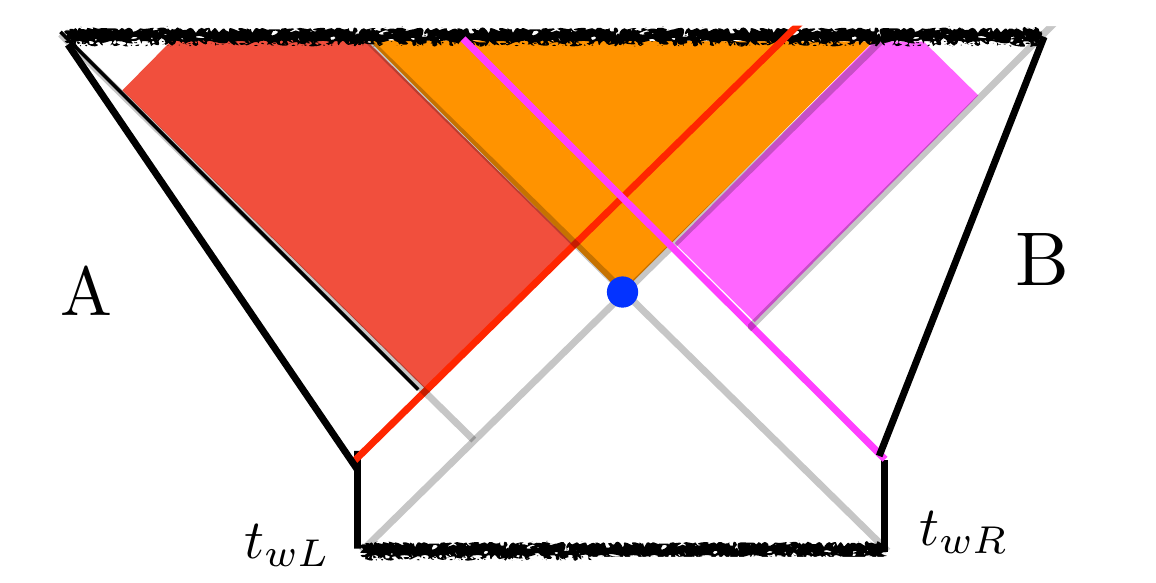}
      \caption{At early time, the entire shared future interior region (orange region) is larger than the post-collision region. }
  \label{Penrose_2shock_3}
  \end{center}
\end{figure}

The total number of healthy gates is given by
\begin{align}
	&\frac{N_{\text{healthy}}^{\text{total}}}{S} = \int_{-\infty}^{+\infty}\frac{2\pi}{\beta}dt\qty(\frac{1}{1+\frac{\delta S_1}{S}e^{\frac{2\pi}{\beta}(-t_{wL}-t)}})\qty(\frac{1}{1+\frac{\delta S_2}{S}e^{\frac{2\pi}{\beta}(t-t_{wR})}})\nonumber\\
	\approx\ &\frac{2\pi}{\beta}(2t_*-\Delta t)\label{total}
\end{align}

On the other hand, the spacetime volume in the entire shared future interior region (orange region in Figure \ref{Penrose_2shock_3}) is given by
\begin{align}
\label{volume_total}
	\frac{V}{\pi r_h l^2} \approx  \frac{2\pi}{\beta}(2t_*-\Delta t)
\end{align}
\eqref{volume_total} agrees with \eqref{total}.

\item{Intermediate time: $t_*\ll \Delta t\ll 2t_*$}

Next, we look at the intermediate regime when $t_*\ll-t_{wL}-t_{wR}\ll2t_*$. Here is the circuit picture. 

\begin{figure}[H] 
 \begin{center}                      
      \includegraphics[width=3.5in]{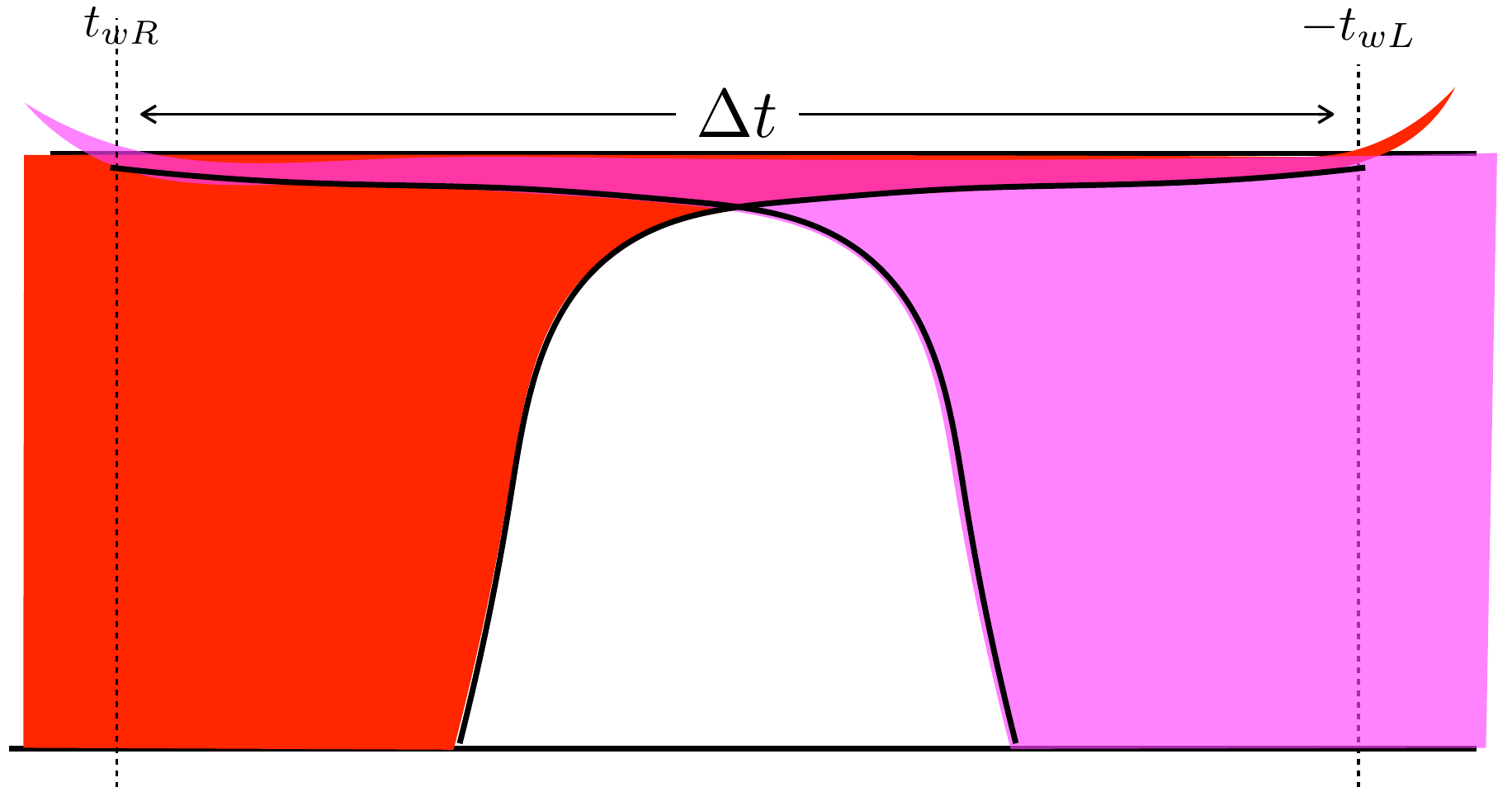}
      \caption{$t_*\ll\Delta t\ll 2t_*$. The number of healthy gates starts to decrease as a function of $\Delta t$.}
  \label{circuit_overlap_2}
  \end{center}
\end{figure}

The number of healthy gates in the overlap region is given by
\begin{align}
\frac{N_{\text{heathy}}}{S} = \ &\int_{t_{wR}}^{-t_{wL}}\frac{2\pi}{\beta}dt\qty(\frac{1}{1+\frac{\delta S_1}{S}e^{\frac{2\pi}{\beta}(-t_{wL}-t)}})\qty(\frac{1}{1+\frac{\delta S_2}{S}e^{\frac{2\pi}{\beta}(t-t_{wR})}})\nonumber\\
\approx\ &\frac{2\pi}{\beta}(2t_*-\Delta t)\qty(1+\frac{\delta S_1\delta S_2e^{\frac{2\pi}{\beta}\Delta t}}{S^2})\label{circuit_intermediate}
\end{align}

We look at the spacetime volume in the intermediate regime. 
\begin{align}
	\frac{V}{\pi\tilde r_h l^2} =\ & \log\frac{2+\frac{2\delta S_1\delta S_2}{S^2}\cosh^2(\frac{\pi}{\beta}\Delta t)}{\frac{2\delta_1\delta_2}{S^2}\cosh^2(\frac{\pi}{\beta}\Delta t)}-\frac{2}{1+\frac{2\delta S_1\delta S_2}{S^2}\cosh^2\frac{\pi}{\beta}\Delta t}\nonumber\\
	\approx \ &\frac{2\pi}{\beta}\qty(2t_*-\Delta t)\nonumber\\
		\frac{V}{\pi r_hl^2} \approx\ & \frac{2\pi}{\beta}(2t_*-\Delta t)\qty(1+\frac{\delta S_1\delta S_2}{2S^2}e^{\frac{2\pi}{\beta}\Delta t})\label{volume_intermediate}
\end{align}

We again see that \eqref{circuit_intermediate} and \eqref{volume_intermediate} agree up to a shift in $\Delta t$.\\

\item{Late time: $\Delta t\gg 2t_*$}

Now we look at the third regime where $\Delta t\gg 2t_*$. From the quantum circuit picture, we no longer expect there to be healthy gates. 
\begin{figure}[H] 
 \begin{center}                      
      \includegraphics[width=4in]{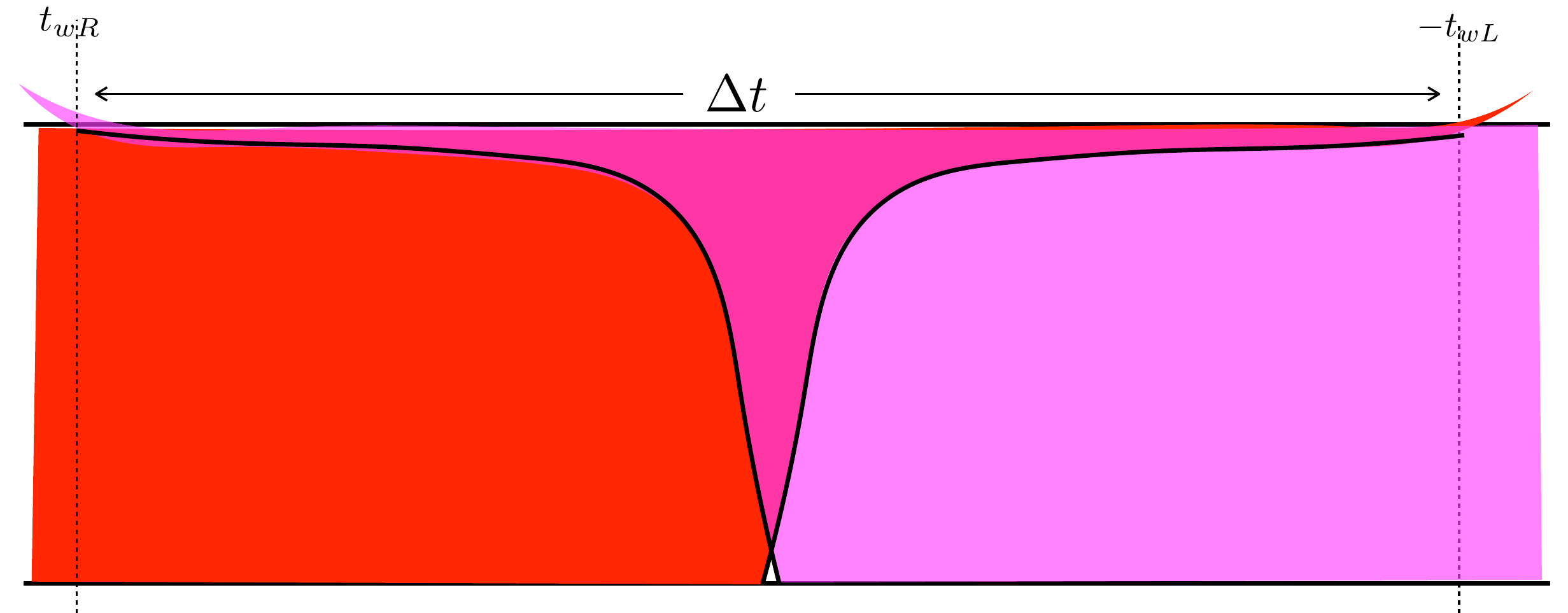}
      \caption{$\Delta t\gg 2t_*$. There are almost no healthy gates. }
  \label{circuit_overlap_3}
  \end{center}
\end{figure}

The epidemic model gives the number of healthy gates
\begin{align*}
\frac{N_{\text{healthy}}}{S}=\ &\int_{t_{wR}}^{-t_{wL}}\frac{2\pi}{\beta}dt\qty(\frac{1}{1+\frac{\delta S_1}{S}e^{\frac{2\pi}{\beta}(-t_{wL}-t)}})\qty(\frac{1}{1+\frac{\delta S_2}{S}e^{\frac{2\pi}{\beta}(t-t_{wR})}})\\
\approx\ &e^{-\frac{2\pi}{\beta}(\Delta t-2t_*)}(\Delta t-2t_*)
\end{align*}

On the gravity side, the post collision region is exponentially small and exponentially close to the singularity. We have
\begin{align*}
	\frac{V}{\pi\tilde r_h l^2} =\ & \log\frac{e^{\frac{\pi}{\beta}(\Delta t-2t_*)}+1}{e^{\frac{\pi}{\beta}(\Delta t-2t_*)}-1}-\frac{2}{e^{\frac{\pi}{\beta}(\Delta t-2t_*)}}\\
	\approx\ &\frac{2}{3}e^{-\frac{3\pi}{\beta}(\Delta t-2t_*)}\\
\frac{V}{\pi r_h l^2} \approx\ &\frac{2}{3}e^{-\frac{2\pi}{\beta}(\Delta t-2t_*)}	
\end{align*}

In this case, the epidemic model and the spacetime volume calculation don't exactly match, but they both decreases exponentially in $\frac{2\pi}{\beta}(\Delta t-2t_*)$.

\end{enumerate}
Here is a plot of the volume of the post collision region and the number of healthy gates as a function of $\Delta t$ in the entire range: $0<\Delta t<3t_*$. We took $\frac{2\pi}{\beta}t_* = 36$. 

\begin{figure}[H] 
 \begin{center}                      
      \includegraphics[width=5in]{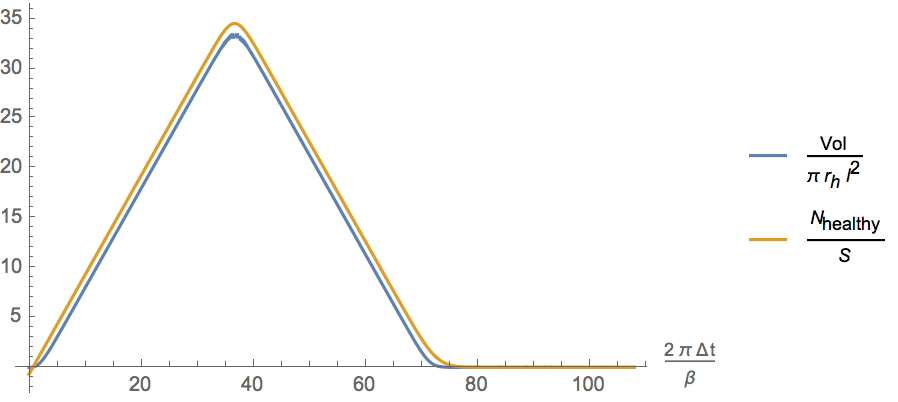}
      \caption{The volume of the post collision region and the number of healthy gates as a function of $\Delta t$. We take $\frac{2\pi}{\beta}t_* = 36$.}
  \label{p2}
  \end{center}
\end{figure}
They differ by an order one constant, corresponding to the gates in one thermal time. One can instead look at their derivative with respect to $\Delta t$:
\begin{figure}[H] 
 \begin{center}                      
      \includegraphics[width=5in]{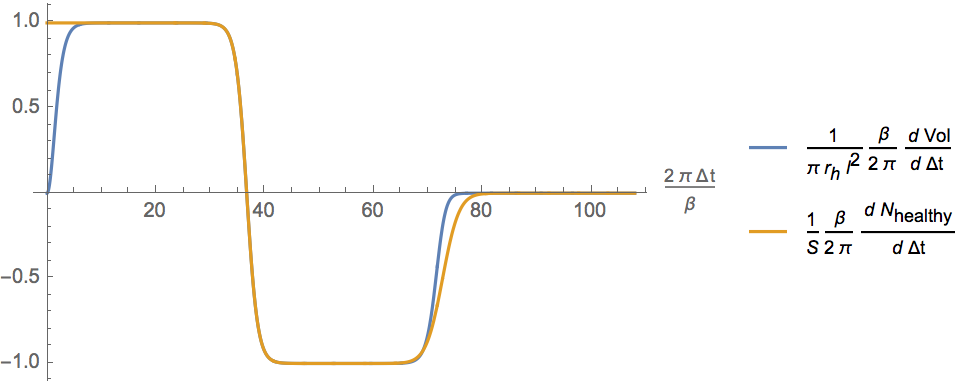}
      \caption{The time derivative of volume of the post collision region and the number of healthy gates as a function of $\Delta t$. We take $\frac{2\pi}{\beta}t_* = 36$.}
  \label{p3}
  \end{center}
\end{figure}
We see good match except for at $\Delta t$ of order thermal time or $\Delta t$ close to $2t_*$. \\

We can also estimate the number of gates contaminated by both epidemics. The ratio of the number of such gates with the number of healthy gates is proportional to the  collision energy. This is a curious fact but we don't have an interpretation.

\subsection{More detailed match of time dependence}

Based on the assumption that the collision in the interior corresponds to the overlap of perturbations in the quantum circuit, we've seen there is a good match between the number of healthy gates in the circuit model and the spacetime volume of the post-collision region. In fact, we can do a more-detailed match. 

We can vary the right time, and see how the spacetime volume in the post-collision region inside the future interior depends on the right time.

\begin{figure}[H] 
 \begin{center}                      
      \includegraphics[width=4in]{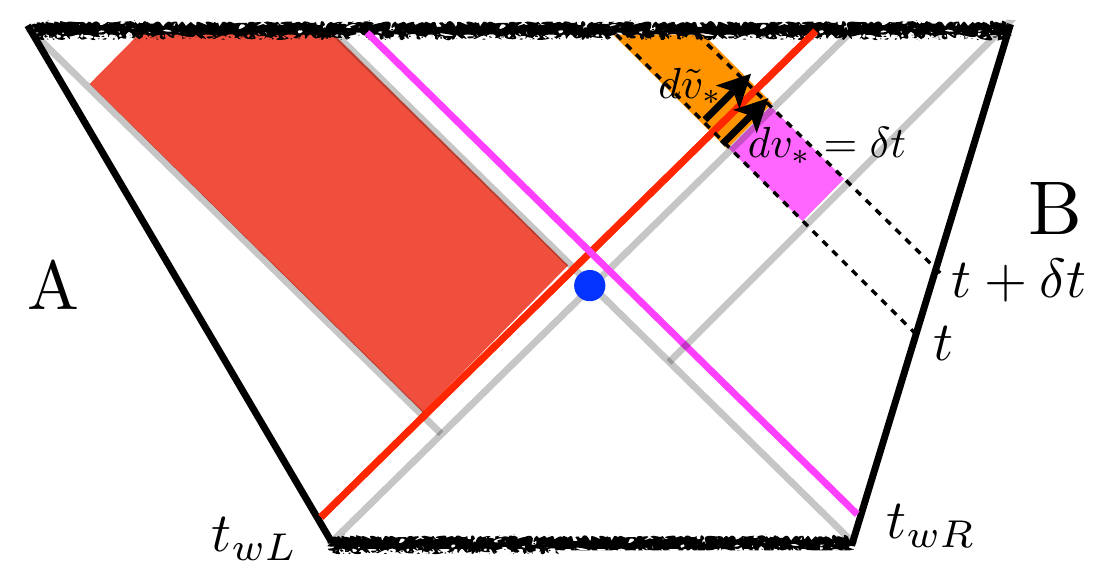}
      \caption{The dependence of the volume of post collision region on the right time.}
  \label{Penrose_2shock_2}
  \end{center}
\end{figure}

We write BTZ metric as $ds^2 = -f(r)dv_*^2+2drdv_*+r^2d\phi^2$. In the post collision region, we replace $f(r)$ by $\tilde f(r)$. 
In Figure \ref{Penrose_2shock_2}, we have
\begin{align*}
	dv_* = \frac{2}{f(r)}dr,\ \ \frac{d\tilde v_*}{dv_*} = \frac{f(r)}{\tilde f(r)}
\end{align*}

\begin{align*}
	\frac{d\text{Vol}_{\text{orange}}}{dt} =\ & \int_0^{r(t)}dr rd\phi \frac{d\tilde v_*}{dv_*}= \pi r(t)^2\frac{r_{h}^2-r(t)^2}{\tilde r_h^2-r(t)^2}
\end{align*}
\begin{align}
	\frac{1}{\pi r_{h}^2}\frac{d\text{Vol}_{\text{orange}}}{dt} =\ & \qty(\frac{1-\frac{\delta S_2}{2S}e^{-\frac{2\pi}{\beta}t_{wR}}e^{\frac{2\pi}{\beta}t}}{1+\frac{\delta S_2}{2S}e^{-\frac{2\pi}{\beta}t_{wR}}e^{\frac{2\pi}{\beta}t}})^2\qty(\frac{1-\qty(\frac{1-\frac{\delta S_2}{2S}e^{-\frac{2\pi}{\beta}t_{wR}}e^{\frac{2\pi}{\beta}t}}{1+\frac{\delta S_2}{2S}e^{-\frac{2\pi}{\beta}t_{wR}}e^{\frac{2\pi}{\beta}t}})^2}{1+\frac{4\delta S_1\delta S_2}{S^2}\cosh^2\frac{\pi}{\beta}(-t_{wL}-t_{wR})-\qty(\frac{1-\frac{\delta S_2}{2S}e^{-\frac{2\pi}{\beta}t_{wR}}e^{\frac{2\pi}{\beta}t}}{1+\frac{\delta S_2}{2S}e^{-\frac{2\pi}{\beta}t_{wR}}e^{\frac{2\pi}{\beta}t}})^2})\nonumber\\
	=\ &\tanh^2(\frac{\pi}{\beta}(t-t_{wR}-t_*))\qty(\frac{1}{1+\cosh^2(\frac{\pi}{\beta}(t_{wR}+t_*-t))\exp((\frac{2\pi}{\beta}(-t_{wL}-t_{wR}-2t_*)))})\nonumber\\
	=\ &\qty(\frac{1}{1+\frac{2\delta S_2}{S}e^{\frac{2\pi}{\beta}(t-t_{wR})}+\mathcal{O}\qty(\qty(\frac{\delta S_2}{S}e^{\frac{2\pi}{\beta}(t-t_{wR})})^2)})\nonumber\\
	&\ \ \ \qty(\frac{1}{1+\frac{\delta S_1}{2S}\exp(\frac{2\pi}{\beta}(-t_{wL}-t))+\frac{1}{2}\exp(\frac{2\pi}{\beta}(\Delta t-2t_*))+\mathcal{O}\qty(\qty(\frac{1}{S}e^{\frac{2\pi}{\beta}(t-t_{wR})})^2)})\label{detail}
\end{align}

We want to compare this with the circuit answer: 
\begin{align}
	\qty(1-\frac{s_{ep}[\frac{2\pi}{\beta}(t-t_{wR})]}{\text{size}_{\text{max}}})\qty(1-\frac{s_{ep}[\frac{2\pi}{\beta}(-t_{wL}-t)]}{\text{size}_{\text{max}}})=\qty(\frac{1}{1+\frac{\delta S_1}{S}e^{\frac{2\pi}{\beta}(t-t_{wR})}})\qty(\frac{1}{1+\frac{\delta S_2}{S}e^{\frac{2\pi}{\beta}(-t_{wL}-t)}})\label{detail_circuit}
\end{align}

We see that these two answers have good match with a shift of $t$ as long as $\frac{1}{S}e^{\frac{2\pi}{\beta}(t-t_{wR})}<1$. This explains the match between the number of healthy gates and the volume of post-collision region. The total volume of post collision region is the integral of \eqref{detail}, while the number of healthy gates is the integral of \eqref{detail_circuit}. 

Here is a plot of \eqref{detail} and \eqref{detail_circuit}, with $\frac{2\pi}{\beta}(t_{wR}+t_*) = 40$, $\frac{2\pi}{\beta}(-t_{wL}-t_*) = 0$. 

\begin{figure}[H] 
 \begin{center}                      
      \includegraphics[width=3.6in]{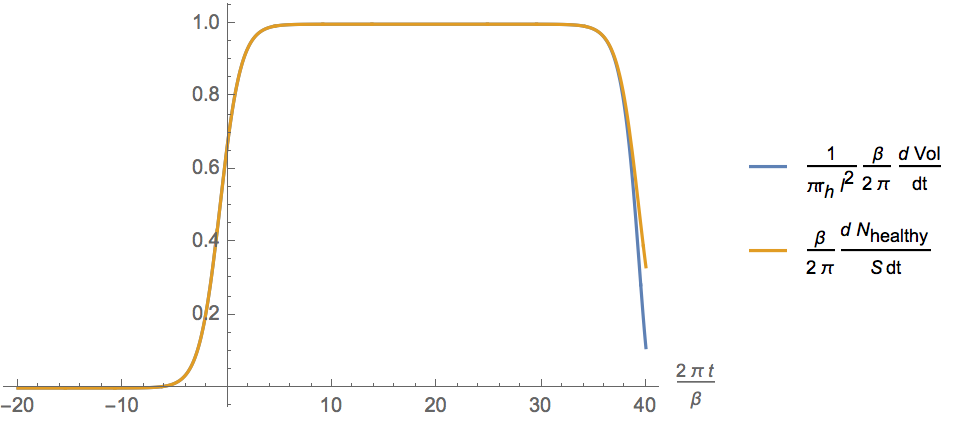}
      \caption{Plot of \eqref{detail} and \eqref{detail_circuit}. We take $\frac{2\pi}{\beta}(t_{wR}+t_*) = 40$, $\frac{2\pi}{\beta}(-t_{wL}-t_*) = 0$.}
  \label{p1}
  \end{center}
\end{figure}

\begin{figure}[H] 
 \begin{center}                      
      \includegraphics[width=3.6in]{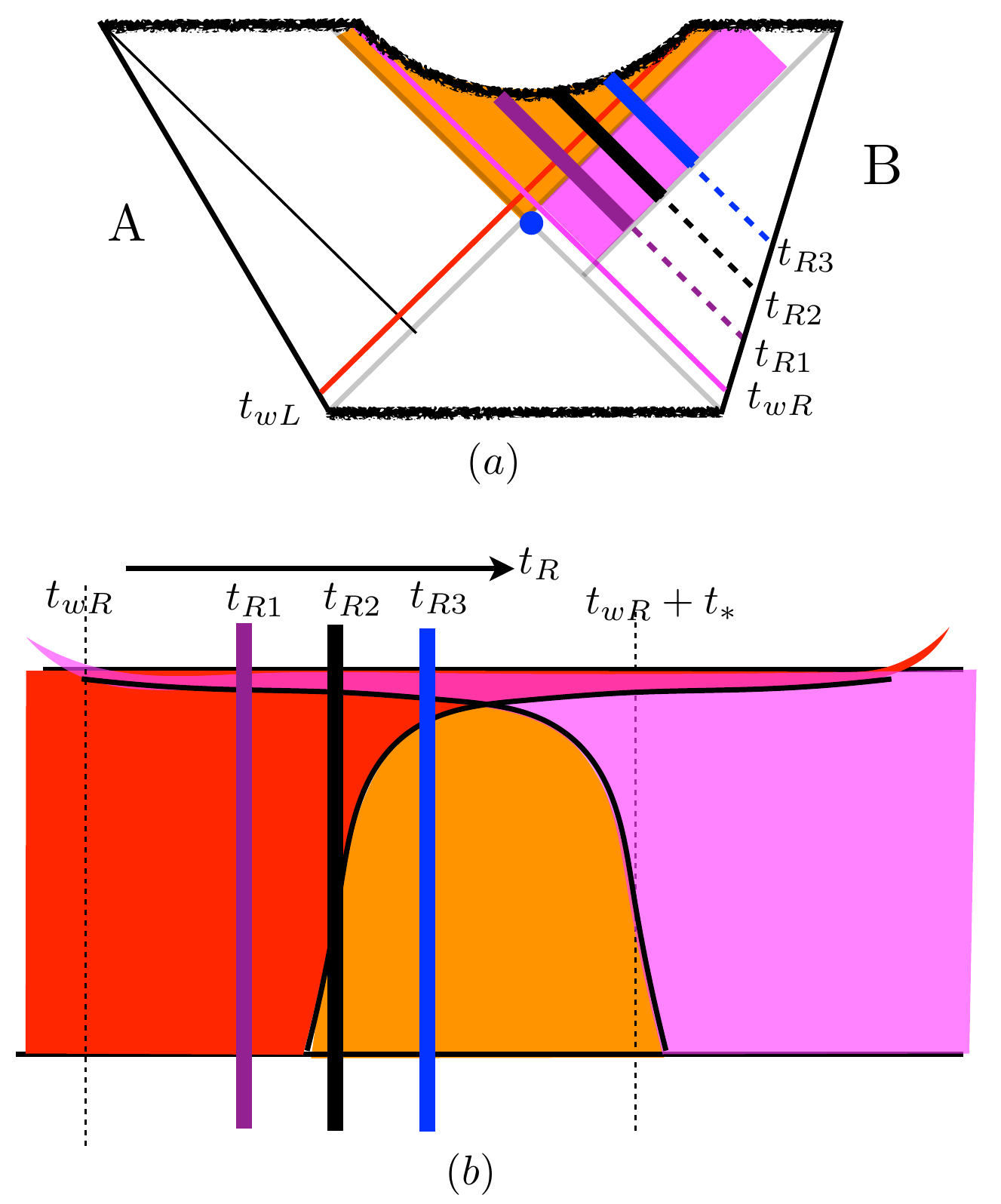}
      \caption{When we increase the right time and see how the volume of the post-collision region changes, it behaves as if we are scanning through the circuit from the left to the right.}
  \label{Penrose_2R}
  \end{center}
\end{figure}

What this computation shows is that, if we focus on how the volume of the post-collision region changes while we increase the right time $t_R$ (Figure \ref{Penrose_2R}(a)), it behaves as if we are scanning through the circuit from left to the right (Figure \ref{Penrose_2R}(b)). 

\section{Conclusion and Discussion}
\label{discussion}

We compared quantum circuit picture with interior geometry, and concluded that the trajectory of an infalling object in the interior near the horizon corresponds to the perturbation in the quantum circuit. 

We also explored the idea that collision in the interior of thermofield double reflects the overlap of perturbations in the quantum circuit, and found good match between the number of healthy gates in the circuit model and the spacetime volume of the post collision region. 

There are many unanswered questions. In this paper we only considered neutral black holes. How do charged black holes fit into the story? For charged black hole, two perturbations falling in from two boundaries at $t_{wL}$ and $t_{wR}$ will collide even when both times are large positive. 

We argued that the collision of objects in the interior corresponds to the overlaps of two perturbations in the circuit. But how to diagnose if the two perturbations overlap or not? The two operators $W_L$ and $W_R$ always commute. 

Our story is only sensitive to the portion of the trajectory of infalling object when it is close to the horizon. We didn't say anything about the singularity. From the picture of neutral black holes, one tends to say that at the singularity the perturbation has smallest size and becomes simple, but this picture doesn't work for charged black holes. 

Another question is how broadly this ``meeting" occurs in general quantum system.\footnote{I thank the referee for asking this question.} We expect the discussion holds when the system has a bulk dual. For more general systems, the meaning of ``meeting" not clear, but we expect that it should be as general as the notion of ER = EPR. One interesting direction to explore is that one can try to get the obejcts out of the black hole after the collision using traversable wormhole type experiment. It was shown in recent development that traversable wormhole phenomena can occur in general quantum systems, though with lower fidelity \cite{Brown:2019hmk}. To study the collision phenomena in that context, one needs to study at least six point functions. We leave this to future work.

\section*{Acknowledgements}
I thank Ahmed Almheiri, Yiming Chen, Adam Levine, Henry Lin, Leonard Susskind, Edward Witten for helpful discussions. I am especially grateful to Juan Maldacena for discussions, detailed comments on the draft, and encouragement. I am supported by the Simons foundation through the It from Qubit Collaboration.

\appendix

\section{Geodesic distance}

We write the BTZ metric as
\begin{align*}
	ds^2 =\ & -f(r)dt^2+\frac{dr^2}{f(r)}+r^2d\phi^2\\
	=\ &-\frac{4l^2 dudv}{(1+uv)^2}+r_h^2\frac{(1-uv)^2}{(1+uv)^2}d\phi^2
\end{align*}
where $f(r) = \frac{r^2-r_h^2}{l^2}$. $u$, $v$ are Kruskal coordinates. $\frac{r}{r_h} = \frac{1-uv}{1+uv}$.

Following \cite{Shenker:2013pqa}, we compute the geodesic distance in BTZ geometry in embedding coordinates. 

\begin{align*}
\cosh\frac{d}{l} = T_1T_1' + T_2 T_2' -X_1 X_1' -X_2 X_2'
\end{align*}
where
\begin{align*}
&T_1 = \frac{v+u}{1+uv} ,\ \ \ T_2 = \frac{1-uv}{1+uv}\cosh\frac{r_h\phi}{l}\\
&X_1 = \frac{v-u}{1+uv} ,\ \ \ X_2 = \frac{1-uv}{1+uv}\sinh\frac{r_h\phi}{l} 
\end{align*}

\subsection{Geodesic distance 1}
\label{geodesic_cal_1}

Alice throws in a perturbation from left side at time $t_w$. When $t_w$ is early enough, the geometry was given in \cite{Dray:1984ha}\cite{Shenker:2013pqa}. There is a shockwave lying close to the horizon $u = 0$. Across the shockwave there is a shift in another Kruskal coordinates $v\longrightarrow v+\alpha$ where $\alpha = \frac{\delta S}{2S}e^{-\frac{2\pi}{\beta}t_w}$, where $S$ is the black hole entropy and $\delta S$ is the increase of the entropy from the perturbation.

 \begin{figure}[H] 
 \begin{center}                      
      \includegraphics[width=3in]{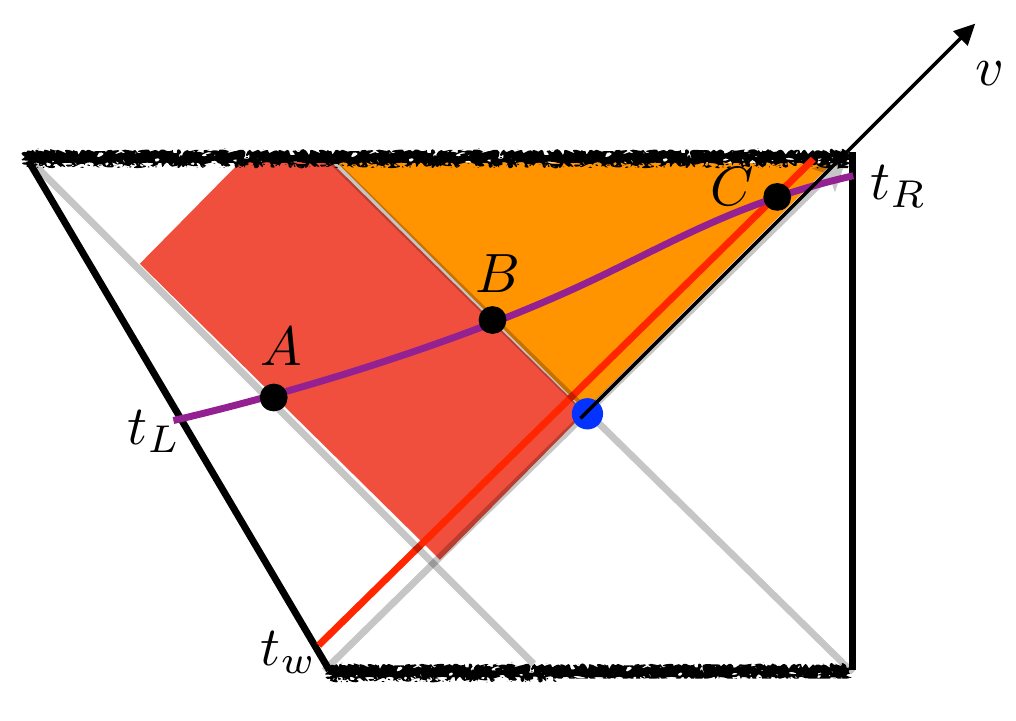}
      \caption{Geodesic length as a function of left time.}
  \label{geodesic_1}
  \end{center}
\end{figure}

We calculate the geodesic length and volume of maximal surface inside and outside Bob's entanglement wedge.

After extremizing the geodesic length between two boundary points anchored at $t_L$ and $t_R$, we get the Kruskal coordinates of the three intermediate points (Figure \ref{geodesic_1}) when we take $t_R\rightarrow\infty$:
\begin{align*}
	&A:(u,v) = \qty(\frac{1}{2}e^{\frac{r_h}{l^2}t_L},0)\\
	&B: (u,v) = \qty(\frac{1}{2e^{-\frac{r_h}{l^2}t_L}+\alpha},\alpha)\\
	&C: (u,v) = \qty(0,\frac{1}{2}\qty(e^{\frac{r_h}{l^2}t_R}+e^{-\frac{r_h}{l^2}t_L}-\alpha))
\end{align*}

We get
\begin{align*}
    \frac{d_{AB}}{l} =\ &\log(1+\alpha e^{t_L})= \ \log(1+e^{\frac{2\pi}{\beta}(t_L-t_w-t_*)})\\
	\frac{d_{BC}}{l} =\ &\frac{2\pi}{\beta}(t_R+t_L)-\log(1+e^{\frac{2\pi}{\beta}(t_L-t_w-t_*)})
\end{align*}

\subsection{Geodesic distance 2}
\label{geodesic_cal_2}

In this section, we fix $t_L\rightarrow +\infty$ and see how length of the part of the geodesic in the orange region depends on the right boundary time. 

\begin{figure}[H] 
 \begin{center}                      
      \includegraphics[width=3in]{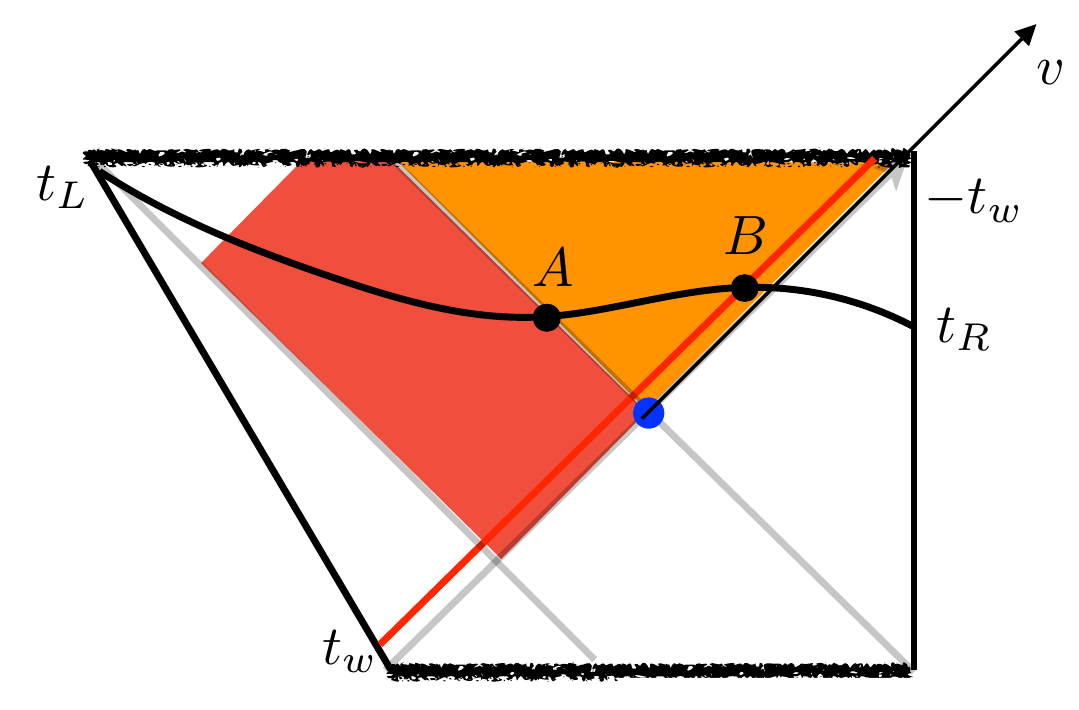}
      \caption{Geodesic length as a function of right time.}
  \label{geodesic_22}
  \end{center}
\end{figure}

In Figure \ref{geodesic_22}, the distance from the left boundary anchored at $t_L$ to point B is given by
\begin{align*}
	\frac{d_{LB}}{l} = \ &\log\frac{2R}{r_H}+\log(\frac{1+e^{\frac{2\pi}{\beta}(t_R+t_L)}+\alpha e^{\frac{2\pi}{\beta}t_L}}{2})\\
\approx\ &\log\frac{R}{r_h}+\frac{2\pi}{\beta}(t_R+t_L)+\log(1+\alpha e^{-\frac{2\pi}{\beta}t_R})
\end{align*}
The distance from the left boundary point to point A is given by
\begin{align*}
	\frac{d_{LA}}{l}\approx \log\frac{2R}{r_h}+\log(1+\alpha e^{\frac{r_h}{l^2}t_L})
\end{align*}
Taking the difference of these two, we get
\begin{align*}
	\frac{d_{AB}}{l}\approx \frac{2\pi}{\beta}(t_R+t_w+t_*)+\log(1+\alpha e^{-\frac{2\pi}{\beta}t_R})
\end{align*}
This is non zero when $t_R>-t_w-t_*$.


\bibliographystyle{apsrev4-1long}
\bibliography{reference}

\end{document}